\input harvmac
\input M5perturbsV1.defs
\noblackbox

\input epsf
  

\input color
  
  
 
\def\journal#1&#2(#3){\unskip, \sl #1\ \bf #2 \rm(19#3) }
\def\andjournal#1&#2(#3){\sl #1~\bf #2 \rm (19#3) }

\def\frac#1#2{{#1\over#2}}

\def\d{\partial}

\def\inbar{\,\vrule height1.5ex width.4pt depth0pt}
\def\IC{\relax\hbox{$\inbar\kern-.3em{\rm C}$}}
\def\IR{\relax{\rm I\kern-.18em R}}
\def\IP{\relax{\rm I\kern-.18em P}}
\def\IZ{\relax{\rm I\kern-.18em Z}}
\def\IE{\relax{\rm I\kern-.18em E}}

\def\dd{\relax{$\delta$\kern-.43em $\delta$}}
\def\dde{\relax{\delta\kern-0.43em \delta}}
%
%

%
\catcode`\@=11
\def\slash#1{\mathord{\mathpalette\c@ncel{#1}}}
\overfullrule=0pt

\def\CC{{\cal C}}

\def\HH{{\cal H}}
\def\II{{\cal I}}

\def\OO{{\cal O}}
\def\PP{{\cal P}}
\def\QQ{{\cal Q}}
\def\RR{{\cal R}}

\def\ZZ{{\cal Z}}

\def\underrel#1\over#2{\mathrel{\mathop{\kern\z@#1}\limits_{#2}}}

\catcode`\@=12


%

\def\det{{\rm det}}

\def\det{{\rm det}}


\def\bG{{\bf \Gamma}}


\def\unlockat{\catcode`\@=11}
\def\lockat{\catcode`\@=12}

\unlockat


\def\newsec#1{\global\advance\secno by1\message{(\the\secno. #1)}
\global\subsecno=0\global\subsubsecno=0\eqnres@t\noindent
{\bf\the\secno. #1}
\writetoca{{\secsym} {#1}}\par\nobreak\medskip\nobreak}
\global\newcount\subsecno \global\subsecno=0
\def\subsec#1{\global\advance\subsecno
by1\message{(\secsym\the\subsecno. #1)}
\ifnum\lastpenalty>9000\else\bigbreak\fi\global\subsubsecno=0
\noindent{\it\secsym\the\subsecno. #1}
\writetoca{\string\quad {\secsym\the\subsecno.} {#1}}
\par\nobreak\medskip\nobreak}
\global\newcount\subsubsecno \global\subsubsecno=0
\def\subsubsec#1{\global\advance\subsubsecno by1
\message{(\secsym\the\subsecno.\the\subsubsecno. #1)}
\ifnum\lastpenalty>9000\else\bigbreak\fi
\noindent\quad{\secsym\the\subsecno.\the\subsubsecno.}{#1}
\writetoca{\string\qquad{\secsym\the\subsecno.\the\subsubsecno.}{#1}}
\par\nobreak\medskip\nobreak}

\def\subsubseclab#1{\DefWarn#1\xdef
#1{\noexpand\hyperref{}{subsubsection}%
{\secsym\the\subsecno.\the\subsubsecno}%
{\secsym\the\subsecno.\the\subsubsecno}}%
\writedef{#1\leftbracket#1}\wrlabeL{#1=#1}}
\def\seclab#1{\xdef #1{\the\secno}\writedef{#1\leftbracket#1}\wrlabeL{#1=#1}}
\def\subseclab#1{\xdef #1{\secsym\the\subsecno}%
\writedef{#1\leftbracket#1}\wrlabeL{#1=#1}}
\lockat


\newcount\figno
\figno=1
\def\fig#1#2#3{
\par\begingroup\parindent=0pt\leftskip=1cm\rightskip=1cm\parindent=0pt
\baselineskip=11pt
\global\advance\figno by 1
\midinsert
\epsfxsize=#3
\centerline{\epsfbox{#2}}
{\bf Fig.\ \the\figno: } #1\par
\endinsert\endgroup\par
}
\def\figlabel#1{\xdef#1{\the\figno}}
\def\encadremath#1{\vbox{\hrule\hbox{\vrule\kern8pt\vbox{\kern8pt
\hbox{$\displaystyle #1$}\kern8pt}
\kern8pt\vrule}\hrule}}
%
%


\font\cmss=cmss10
\font\cmsss=cmss10 at 7pt
\def\rlx{\relax\leavevmode}
\def\inbar{\vrule height1.5ex width.4pt depth0pt}
\def\IC{\relax\,\hbox{$\inbar\kern-.3em{\rm C}$}}
\def\IN{\relax{\rm I\kern-.18em N}}
\def\IP{\relax{\rm I\kern-.18em P}}
\def\ZZ{\rlx\leavevmode\ifmmode\mathchoice{\hbox{\cmss Z\kern-.4em Z}}
 {\hbox{\cmss Z\kern-.4em Z}}{\lower.9pt\hbox{\cmsss Z\kern-.36em Z}}
 {\lower1.2pt\hbox{\cmsss Z\kern-.36em Z}}\else{\cmss Z\kern-.4em
 Z}\fi}
\def\IZ{\relax\ifmmode\mathchoice
{\hbox{\cmss Z\kern-.4em Z}}{\hbox{\cmss Z\kern-.4em Z}}
{\lower.9pt\hbox{\cmsss Z\kern-.4em Z}}
{\lower1.2pt\hbox{\cmsss Z\kern-.4em Z}}\else{\cmss Z\kern-.4em
Z}\fi}
\def\IZ{\relax\ifmmode\mathchoice
{\hbox{\cmss Z\kern-.4em Z}}{\hbox{\cmss Z\kern-.4em Z}}
{\lower.9pt\hbox{\cmsss Z\kern-.4em Z}}
{\lower1.2pt\hbox{\cmsss Z\kern-.4em Z}}\else{\cmss Z\kern-.4em
Z}\fi}

\def\narrowplus{\kern -.04truein + \kern -.03truein}
\def\narrowminus{- \kern -.04truein}
\def\narrowminussub{\kern -.02truein - \kern -.01truein}

\def\IZ{\relax\ifmmode\mathchoice
{\hbox{\cmss Z\kern-.4em Z}}{\hbox{\cmss Z\kern-.4em Z}}
{\lower.9pt\hbox{\cmsss Z\kern-.4em Z}}
{\lower1.2pt\hbox{\cmsss Z\kern-.4em Z}}\else{\cmss Z\kern-.4em
Z}\fi}
\def\IB{\relax{\rm I\kern-.18em B}}
\def\IC{{\relax\hbox{$\inbar\kern-.3em{\rm C}$}}}
\def\ID{\relax{\rm I\kern-.18em D}}
\def\IE{\relax{\rm I\kern-.18em E}}
\def\IF{\relax{\rm I\kern-.18em F}}
\def\IG{\relax\hbox{$\inbar\kern-.3em{\rm G}$}}
\def\IGa{\relax\hbox{${\rm I}\kern-.18em\Gamma$}}
\def\IH{\relax{\rm I\kern-.18em H}}
\def\II{\relax{\rm I\kern-.18em I}}
\def\IK{\relax{\rm I\kern-.18em K}}
\def\IP{\relax{\rm I\kern-.18em P}}

\font\cmss=cmss10 \font\cmsss=cmss10 at 7pt
\def\IR{\relax{\rm I\kern-.18em R}}

\def\Fslash{F\llap{/}}
\def\dslash{\partial\llap{/}}
%

%
%
\def\eqnn#1{\xdef #1{(\secsym\the\meqno)}\writedef{#1\leftbracket#1}%
\global\advance\meqno by1\wrlabeL#1}
\def\eqna#1{\xdef #1##1{\hbox{$(\secsym\the\meqno##1)$}}
\writedef{#1\numbersign1\leftbracket#1{\numbersign1}}%
\global\advance\meqno by1\wrlabeL{#1$\{\}$}}
\def\eqn#1#2{\xdef #1{(\secsym\the\meqno)}\writedef{#1\leftbracket#1}%
\global\advance\meqno by1$$#2\eqno#1\eqlabeL#1$$}


\def\boxit#1{\vbox{\hrule\hbox{\vrule\kern8pt
\vbox{\hbox{\kern8pt}\hbox{\vbox{#1}}\hbox{\kern8pt}}
\kern8pt\vrule}\hrule}}
\def\mathboxit#1{\vbox{\hrule\hbox{\vrule\kern5pt\vbox{\kern5pt
\hbox{$\displaystyle #1$}\kern5pt}\kern5pt\vrule}\hrule}}


\lref\Izquierdo{
  J.~M.~Izquierdo, N.~D.~Lambert, G.~Papadopoulos and P.~K.~Townsend,
  ``Dyonic membranes,''
  Nucl.\ Phys.\ B {\bf 460} (1996) 560
  [hep-th/9508177].
}

\lref\EmparanCS{
  R.~Emparan, T.~Harmark, V.~Niarchos and N.~A.~Obers,
  ``World-Volume Effective Theory for Higher-Dimensional Black Holes,''
Phys.\ Rev.\ Lett.\  {\bf 102}, 191301 (2009).
[arXiv:0902.0427 [hep-th]].
}

\lref\EmparanHG{
  R.~Emparan, T.~Harmark, V.~Niarchos and N.~A.~Obers,
  ``Blackfolds in Supergravity and String Theory,''
JHEP {\bf 1108}, 154 (2011).
[arXiv:1106.4428 [hep-th]].
}

\lref\EmparanAT{
  R.~Emparan, T.~Harmark, V.~Niarchos and N.~A.~Obers,
  ``Essentials of Blackfold Dynamics,''
  JHEP {\bf 1003} (2010) 063
  [arXiv:0910.1601 [hep-th]].
}

\lref\CaldarelliXZ{
  M.~M.~Caldarelli, R.~Emparan and B.~Van Pol,
  ``Higher-dimensional Rotating Charged Black Holes,''
JHEP {\bf 1104}, 013 (2011).
[arXiv:1012.4517 [hep-th]].
}

\lref\HarmarkFF{
  T.~Harmark,
  ``Open branes in space-time noncommutative little string theory,''
Nucl.\ Phys.\ B {\bf 593}, 76 (2001).
[hep-th/0007147].
}

\lref\HarmarkRB{
  T.~Harmark and N.~A.~Obers,
  ``Phase structure of noncommutative field theories and spinning brane bound states,''
JHEP {\bf 0003}, 024 (2000).
[hep-th/9911169].
}

\lref\GrignaniXM{
  G.~Grignani, T.~Harmark, A.~Marini, N.~A.~Obers and M.~Orselli,
  ``Heating up the BIon,''
JHEP {\bf 1106}, 058 (2011).
[arXiv:1012.1494 [hep-th]].
}

\lref\GrignaniXMM{
  G.~Grignani, T.~Harmark, A.~Marini, N.~A.~Obers and M.~Orselli,
  ``Thermodynamics of the hot BIon,''
  Nucl.\ Phys.\ B {\bf 851} (2011) 462
  [arXiv:1101.1297 [hep-th]].
}

\lref\PastiGX{
  P.~Pasti, D.~P.~Sorokin and M.~Tonin,
  ``Covariant action for a D = 11 five-brane with the chiral field,''
Phys.\ Lett.\ B {\bf 398}, 41 (1997).
[hep-th/9701037].
}

\lref\BandosUI{
  I.~A.~Bandos, K.~Lechner, A.~Nurmagambetov, P.~Pasti, D.~P.~Sorokin and M.~Tonin,
  ``Covariant action for the superfive-brane of M theory,''
Phys.\ Rev.\ Lett.\  {\bf 78}, 4332 (1997).
[hep-th/9701149].
}

\lref\PerryMK{
  M.~Perry and J.~H.~Schwarz,
  ``Interacting chiral gauge fields in six-dimensions and Born-Infeld theory,''
Nucl.\ Phys.\ B {\bf 489}, 47 (1997).
[hep-th/9611065].
}

\lref\CampsHW{
  J.~Camps and R.~Emparan,
  ``Derivation of the blackfold effective theory,''
JHEP {\bf 1203}, 038 (2012).
[arXiv:1201.3506 [hep-th]].
}

\lref\Simon{
  J.~Simon,
  ``Brane Effective Actions, Kappa-Symmetry and Applications,''
  Living Rev.\ Rel.\  {\bf 5} (2012) 3
  [arXiv:1110.2422 [hep-th]].
}

\lref\NiarchosCY{
  V.~Niarchos and K.~Siampos,
  ``Entropy of the self-dual string soliton,''
JHEP {\bf 1207}, 134 (2012).
[arXiv:1206.2935 [hep-th]].
}

\lref\LuNU{
  H.~Lu, C.~N.~Pope, J.~Rahmfeld and ,
  ``A Construction of Killing spinors on S**n,''
J.\ Math.\ Phys.\  {\bf 40}, 4518 (1999).
[hep-th/9805151].
}

\lref\EmparanWM{
  R.~Emparan, T.~Harmark, V.~Niarchos, N.~A.~Obers and M.~J.~Rodriguez,
  ``The Phase Structure of Higher-Dimensional Black Rings and Black Holes,''
JHEP {\bf 0710}, 110 (2007).
[arXiv:0708.2181 [hep-th]].
}

\lref\GorbonosUC{
  D.~Gorbonos and B.~Kol,
  ``A Dialogue of multipoles: Matched asymptotic expansion for caged black holes,''
JHEP {\bf 0406}, 053 (2004).
[hep-th/0406002].
}

\lref\NiarchosPN{
  V.~Niarchos and K.~Siampos,
  ``M2-M5 blackfold funnels,''
JHEP {\bf 1206}, 175 (2012).
[arXiv:1205.1535 [hep-th]].
}

\lref\GauntlettFZ{
  J.~P.~Gauntlett and S.~Pakis,
  ``The Geometry of D = 11 killing spinors,''
JHEP {\bf 0304}, 039 (2003).
[hep-th/0212008].
}

\lref\BergshoeffJN{
  E.~Bergshoeff, D.~S.~Berman, J.~P.~van der Schaar and P.~Sundell,
  ``A Noncommutative M theory five-brane,''
Nucl.\ Phys.\ B {\bf 590}, 173 (2000).
[hep-th/0005026].
}

\lref\GopakumarEP{
  R.~Gopakumar, S.~Minwalla, N.~Seiberg and A.~Strominger,
  ``(OM) theory in diverse dimensions,''
JHEP {\bf 0008}, 008 (2000).
[hep-th/0006062].
}

\lref\CampsBR{
  J.~Camps, R.~Emparan and N.~Haddad,
  ``Black Brane Viscosity and the Gregory-Laflamme Instability,''
JHEP {\bf 1005}, 042 (2010).
[arXiv:1003.3636 [hep-th]].
}

\lref\FluxBlackfolds{
 J.~ Armas, J.~ Gath, V.~Niarchos, N.~A.~Obers and A.~V.~Pedersen,
  to appear.
}

\lref\NiarchosIA{
  V.~Niarchos and K.~Siampos,
  ``The black M2-M5 ring intersection spins,''
PoS Corfu {\bf 2012}, 088 (2013).
[arXiv:1302.0854 [hep-th]].
}

\lref\BhattacharyyaJC{
  S.~Bhattacharyya, V.~E.~Hubeny, S.~Minwalla and M.~Rangamani,
  ``Nonlinear Fluid Dynamics from Gravity,''
JHEP {\bf 0802}, 045 (2008).
[arXiv:0712.2456 [hep-th]].
}

\lref\GathQYA{
  J.~Gath and A.~V.~Pedersen,
  ``Viscous Asymptotically Flat Reissner-Nordstr\"om Black Branes,''
[arXiv:1302.5480 [hep-th]].
}

\lref\EmparanUX{
  R.~Emparan, D.~Mateos and P.~K.~Townsend,
  ``Supergravity supertubes,''
JHEP {\bf 0107}, 011 (2001).
[hep-th/0106012].
}

\lref\LuninMJ{
  O.~Lunin,
  ``Strings ending on branes from supergravity,''
JHEP {\bf 0709}, 093 (2007).
[arXiv:0706.3396 [hep-th]].
}

\lref\LuninTF{
  O.~Lunin,
  ``Brane webs and 1/4-BPS geometries,''
JHEP {\bf 0809}, 028 (2008).
[arXiv:0802.0735 [hep-th]].
}

\lref\TseytlinDJ{
  A.~A.~Tseytlin,
  ``Born-Infeld action, supersymmetry and string theory,''
In *Shifman, M.A. (ed.): The many faces of the superworld* 417-452.
[hep-th/9908105].
}

\lref\MyersBW{
  R.~C.~Myers,
  ``NonAbelian phenomena on D branes,''
Class.\ Quant.\ Grav.\  {\bf 20}, S347 (2003).
[hep-th/0303072].
}

\lref\EmparanILA{
  R.~Emparan, V.~E.~Hubeny and M.~Rangamani,
  ``Effective hydrodynamics of black D3-branes,''
JHEP {\bf 1306}, 035 (2013).
[arXiv:1303.3563 [hep-th]].
}

\lref\GrignaniEWA{
  G.~Grignani, T.~Harmark, A.~Marini and M.~Orselli,
  ``Thermal DBI action for the D3-brane at weak and strong coupling,''
[arXiv:1311.3834 [hep-th]].
}

\lref\GrignaniIW{
  G.~Grignani, T.~Harmark, A.~Marini, N.~A.~Obers and M.~Orselli,
  ``Thermal string probes in AdS and finite temperature Wilson loops,''
JHEP {\bf 1206}, 144 (2012).
[arXiv:1201.4862 [hep-th]].
}

\lref\ArmasBK{
  J.~Armas, T.~Harmark, N.~A.~Obers, M.~Orselli and A.~V.~Pedersen,
  ``Thermal Giant Gravitons,''
JHEP {\bf 1211}, 123 (2012).
[arXiv:1207.2789 [hep-th]].
}

\lref\EmparanVD{
  R.~Emparan, T.~Harmark, V.~Niarchos and N.~A.~Obers,
  ``New Horizons for Black Holes and Branes,''
JHEP {\bf 1004}, 046 (2010).
[arXiv:0912.2352 [hep-th]].
}

\lref\CampsHB{
  J.~Camps, R.~Emparan, P.~Figueras, S.~Giusto and A.~Saxena,
  ``Black Rings in Taub-NUT and D0-D6 interactions,''
JHEP {\bf 0902}, 021 (2009).
[arXiv:0811.2088 [hep-th]].
}

\lref\ArmasHZ{
  J.~Armas and N.~A.~Obers,
  ``Blackfolds in (Anti)-de Sitter Backgrounds,''
Phys.\ Rev.\ D {\bf 83}, 084039 (2011).
[arXiv:1012.5081 [hep-th]].
}

\lref\ArmasOTA{
  J.~Armas, N.~A.~Obers and A.~V.~Pedersen,
  ``Null-Wave Giant Gravitons from Thermal Spinning Brane Probes,''
JHEP {\bf 1310}, 109 (2013).
[arXiv:1306.2633 [hep-th]].
}

\lref\ArmasAKA{
  J.~Armas, J.~Gath and N.~A.~Obers,
  ``Electroelasticity of Charged Black Branes,''
JHEP {\bf 1310}, 035 (2013).
[arXiv:1307.0504 [hep-th]].
}

\lref\PastiVS{
  P.~Pasti, D.~P.~Sorokin and M.~Tonin,
  ``On Lorentz invariant actions for chiral p forms,''
Phys.\ Rev.\ D {\bf 55}, 6292 (1997).
[hep-th/9611100].
}

\lref\MartucciDN{
  L.~Martucci,
  ``Electrified branes,''
JHEP {\bf 1202}, 097 (2012).
[arXiv:1110.0627 [hep-th]].
}

\lref\GauntlettDI{
  J.~P.~Gauntlett,
  ``Branes, calibrations and supergravity,''
[hep-th/0305074].
}

\lref\CallanKY{
  C.~G.~Callan, Jr., J.~A.~Harvey and A.~Strominger,
  ``Worldbrane actions for string solitons,''
Nucl.\ Phys.\ B {\bf 367}, 60 (1991).
}

\lref\GibbonsSV{
  G.~W.~Gibbons and P.~K.~Townsend,
  ``Vacuum interpolation in supergravity via super p-branes,''
Phys.\ Rev.\ Lett.\  {\bf 71}, 3754 (1993).
[hep-th/9307049].
}

\lref\CaldarelliHY{
  M.~M.~Caldarelli, J.~Camps, B.~Gout\'eraux and K.~Skenderis,
  ``AdS/Ricci-flat correspondence and the Gregory-Laflamme instability,''
Phys.\ Rev.\ D {\bf 87}, no. 6, 061502 (2013).
[arXiv:1211.2815 [hep-th]].
}

\lref\CaldarelliAAA{
  M.~M.~Caldarelli, J.~Camps, B.~Gout\'eraux and K.~Skenderis,
  ``AdS/Ricci-flat correspondence,''
[arXiv:1312.7874 [hep-th]].
}



\rightline{CCQCN-2014-23}
\rightline{CCTP-2014-27} 
\vskip 15pt
\Title{
}
{\vbox{\centerline{Supersymmetric Perturbations of the M5 brane}
}}
\bigskip
\centerline{Vasilis Niarchos
}
\bigskip
\centerline{{\it Crete Center for Theoretical Physics}}
\centerline{{\it \& Crete Center for Quantum Complexity and Nanotechnology}}
\medskip
\centerline{\it Department of Physics, University of Crete, 71303, Greece}
\bigskip
\centerline{niarchos@physics.uoc.gr}
\bigskip\bigskip\bigskip
\centerline{\bf Abstract}
\bigskip

\noindent 
We study long-wavelength supersymmetric deformations of brane solutions in supergravity 
using an extension of previous ideas within the general scheme of the blackfold approach. 
As a concrete example, we consider long-wavelength perturbations of the planar M2-M5 bound state solution
in eleven-dimensional supergravity. We propose a specific ansatz for the first order deformation of the 
supergravity fields and explore how this deformation perturbs the Killing spinor equations.
We find that a special part of these equations gives a projection equation on the Killing spinors 
that has the same structure as the $\kappa$-symmetry condition of the abelian M5 brane theory. 
Requiring a match between supergravity and gauge theory implies a specific non-linear gauge-gravity 
map between the bosonic fields of the abelian M5 brane theory and the gravity-induced fluid-like 
degrees of freedom of the blackfold equations that control the perturbative gravity solution. 
This observation sheds new light on the SUGRA/DBI correspondence.

\vfill
\bigskip
\Date{}


\listtoc
\writetoc
\writedefs

\newsec{Introduction}

\subsec{The forest}

D-/M-branes in string/M-theory have a rich space of supersymmetric configurations.
This space is parameterized by a set of discrete parameters (abstractly, the number $N$ of branes),
a set of continuous parameters (the moduli), which are vacuum expectation values of gauge-invariant
operators, and the field profiles that define the bulk (closed string) background on which the brane propagates. 
The expectation values of the moduli may or may not break the underlying gauge symmetry of the brane. 
For concreteness, we will focus on the subspace of supersymmetric (bosonic)
configurations that do not break the gauge symmetry (in a standard system like the system of 
$N$ D3 branes this would be the origin of the Coulomb branch). 

Tracing the properties of this space across $N$ requires a detailed understanding of the theory
that resides on the brane. For D-branes in string theory a configuration of the brane (supersymmetric or
not) is determined by an exact solution of an open string (field) theory, which, in general, is hard to obtain 
directly. The common approach to this problem is to identify first an exact solution (a special point 
$P$ in configuration space) and work in its vicinity by setting up an effective
field theory description of long-wavength deformations. $P$ is usually a point with enhanced symmetry 
where the full open string equations of motion can be solved exactly. The general assumption in all practical 
applications of this strategy is that 
solutions of the leading order effective field theory are approximate representations of an exact configuration 
a finite distance away from $P$.

It is best known how to implement this stategy in the abelian case of a single D-brane in string theory. 
In a standard derivation (see \TseytlinDJ\ for a review) $P$ is a brane configuration with a flat 
worldvolume and a constant gauge field strength $F_2$. The resulting effective field theory, 
which is the well-known abelian Dirac-Born-Infeld (DBI) theory, is determined from a computation of the disk 
worldsheet partition function or a computation of an infinite set of tree-level scattering amplitudes. 
The extension to curved backgrounds and the inclusion of general
couplings with the background fields is known. However, the understanding of the 
non-abelian extension of this theory $(N\geq 2)$ is more rudimentary (we refer the reader to the review
\MyersBW\ and references therein).

Configurations that preserve some amount of supersymmetry are special and obey additional conditions. 
These conditions are nicely packaged in a single projection equation
\eqn\introaa{
\bG_\kappa \, \epsilon = \epsilon~, ~~ \bG^2_\kappa = {\bf 1}
}
that involves the spinorial supersymmetry transformation parameter $\epsilon$ \Simon. 
$\bG_\kappa$ is the matrix that controls the 
$\kappa$-symmetry transformations of the brane theory. The specifics of this matrix depend on the 
profile of the brane and background fields. The solutions of this equation involve
$(i)$ a set of first order differential constraints on the bosonic fields of the brane effective theory, and
$(ii)$ a set of differential and algebraic constraints on $\epsilon$ that determine the amount of the 
preserved supersymmetry.

It is of interest to understand how these structures evolve as we increase the number of branes $N$.
Parts of the configuration space can often be traced from the one extreme at $N=1$ to the
other at $N \to \infty$. For instance, configurations at finite $N$ that involve only the abelian (diagonal $U(1)$, 
center-of-mass) degrees of freedom are always solutions of the same abelian effective DBI theory 
(up to overall constants). The $N\to \infty$ regime is particularly interesting, because it is the regime
where configurations are described by classical solutions of the bulk supergravity theory. There are 
many instances in the literature where a direct correspondence is observed between solutions of the abelian
DBI theory and brane solutions of the bulk supergravity equations of motion (a SUGRA/DBI correspondence).
An indicative list of examples includes \refs{\EmparanUX,\LuninMJ,\LuninTF}. Connections between 
$\kappa$-symmetry and calibrations in brane theory and supergravity are also relevant for this aspect (for 
recent related work see \MartucciDN; for an instructive review see \GauntlettDI). To the best of our knowledge,
there is currently no systematic general understanding of this correspondence and part of the motivation of the 
present work is to provide one.

Such comparisons between the low-$N$ and large-$N$ descriptions require a continuation of the 
DBI philosophy to large-$N$, namely the notion of a theory of long-wavelength deformations of 
brane solutions in supergravity. In recent years it has been proposed \refs{\EmparanCS,\EmparanAT} 
that such a theory can be set up within a scheme of matched asymptotic expansions 
(see \GorbonosUC\ for a detailed discussion of such expansions in the context of caged black holes). 
Once the zeroth order solution is identified, the constant parameters that control it ($e.g.$
charge densities, rotation parameters, $etc.$) are promoted to slowly-varying fields of the worldvolume
coordinates and the gravity equations are solved perturbatively 
in a derivative expansion scheme. In the process one discovers that the slowly-varying fields are degrees 
of freedom of a gravity-induced effective worldvolume theory (coined blackfold theory \EmparanCS)
that involves a fluid on a dynamical hypersurface. The hypersurface acts in many ways as
a holographic screen; it is naturally located in a region far from the black hole horizon and the theory 
on it is conjectured to control the perturbative solution in the bulk. 

This approach can be employed in diverse contexts, $e.g.$ zero or finite temperature, neutral or 
charged branes, asymptotically flat or asymptotically non-flat backgrounds. 
For finite-temperature AdS black branes the same process leads naturally to the fluid-gravity correspondence 
\BhattacharyyaJC, where one recovers an effective fluid on a surface with fixed geometry. The latter 
is related holographically with the fluid description of a dual strongly coupled quantum field theory. 
Until now the blackfold approach has been applied successfully in gravity to provide evidence for new
black holes solutions with exotic horizon geometries 
\refs{\EmparanWM\CampsHB\EmparanVD\ArmasHZ\CaldarelliXZ-\EmparanHG}, 
in the AdS/CFT correspondence \refs{\GrignaniIW,\ArmasBK}, in string/M-theory 
\refs{\GrignaniXM\GrignaniXMM\NiarchosPN\NiarchosCY\NiarchosIA\ArmasOTA\ArmasAKA-\GrignaniEWA}. 
A recent discussion about the relation of blackfolds, the fluid-gravity correspondence
and the membrane paradigm appeared in \EmparanILA. A different interesting direction has been pursued
in \refs{\CaldarelliHY,\CaldarelliAAA}.

For asymptotically flat neutral black brane solutions in Einstein gravity Refs.\ \refs{\CampsBR,\CampsHW} 
have shown that the leading order hydrodynamical equations of blackfold theory guarantee the existence of a 
regular first order corrected solution of the Einstein equations. Regularity refers here to the solution outside the 
black hole horizon. This result has been partially extended to electrically charged black brane solutions of 
Einstein-Maxwell gravity in \GathQYA. 
The generic brane configuration in theories with bulk gauge fields ($e.g.$ in supergravity) 
involves complex anisotropic fluids with multiple conserved higher-differential-form charge currents. 
Such descriptions are ubiquitous in the discussion of brane bound states. 
The elastic and fluid-dynamical properties of these systems are less understood.
More importantly, the derivation of the first order corrected supergravity solutions in these more general cases 
remains a largely open problem. 

Supersymmetric solutions provide a fruitful arena for this problem. 
As a first step in this direction we consider the formulation of supersymmetric perturbations of brane bound
states within the blackfold expansion scheme. We propose a specific ansatz for the first order deformation of the 
supergravity fields and demonstrate that a special part of the supergravity Killing spinor equations gives rise to a 
projection equation that has the same structure as the $\kappa$-symmetry equation \introaa\ of the non-
gravitational abelian DBI theory. This observation suggests a concrete gauge-gravity map between the 
degrees of freedom of the DBI description and degrees of freedom derived directly from gravity.
As a result, this map adds a new element to the SUGRA/DBI correspondence and opens the road to a deeper 
understanding of the nature and structure of the dynamical equations of the blackfold effective theory.

As a final comment, we note that there is an old proposal (first applied to string theory in \CallanKY) that 
identifies the abelian part of the D-/M-brane degrees of freedom (transverse scalars, gauge fields) in supergravity
as collective coordinates associated to large gauge transformations. Although this approach shares 
some qualitative similarities with the blackfold description, the two are significantly different. For instance,
the old approach of \CallanKY\ (for a review see \Simon) identifies from supergravity worldvolume abelian gauge 
fields. In contrast, we will see that the blackfold approach identifies naturally, with a particular non-linear
rewriting that we analyze in this paper, worldvolume abelian gauge field strengths.\foot{For brane configurations 
in supergravity interpolating
between the asymptotic Minkowski space and a near-horizon AdS space, the abelian nature of the brane
effective theory suggests that this theory is a supersingleton field theory. This fact was recognized early on
in \GibbonsSV. The new element that we add to this story is the proposal that this supersingleton field theory is
blackfold theory. It would be interesting to explore this connection further.} 
Moreover, the blackfold effective theory encodes in a rather straightforward manner the full non-linear nature of 
the DBI action. This is hard to achieve with the techniques of \CallanKY.

\subsec{The tree} 

For concreteness, in this paper we will investigate the above construction in the context of a very specific
brane configuration in the eleven-dimensional supergravity description of M-theory. Following closely
the logic of the string theory derivation of the DBI action \TseytlinDJ\ we begin with the configuration of an 
M5 brane in the presence of a constant worldvolume three-form flux. The specifics of this solution and 
the details of the abelian M5 brane theory proposed in \refs{\PastiVS\PastiGX-\BandosUI} are summarized 
in section \Mdbi. In section \Msugra\ we 
move to the supergravity regime and recall the details of a corresponding exactly known supergravity 
solution that represents a planar M2-M5 bound state. Both gauge theory and gravity solutions are 1/2-BPS and 
the analysis of the Killing spinor equations in supergravity in this simple uniform case reveals immediately a 
natural connection with the $\kappa$-symmetry equation of the abelian M5 brane theory. Part of this 
connection is a specific map between the uniform three-form flux of the abelian M5 brane theory
and a parameter that controls the M2 brane charge in the supergravity solution. 

An ansatz for a general extremal long-wavelength deformation of the seed planar M2-M5 solution in 
supergravity is proposed in section \generalansatz. The proposal extends the treatment of neutral
black branes in \refs{\CampsBR,\CampsHW} to a setting of an extremal two-charge black brane solution.
We summarize the basic assumptions and salient features of the approach and list the relevant bosonic
blackfold equations.

The insertion of this ansatz to the supergravity Killing spinor equations gives a lengthy set of complicated
equations. In sections \intrinsic, \extrinsic, which contain the main results of the paper, we notice that
there are drastic simplifications if we focus on a specific part of the equations, and that this part has the 
same structure as the $\kappa$-symmetry equations of the abelian M5 brane theory. Having considered 
a general inhomogeneous configuration of the brane solution, the correspondence between the 
gravity-induced and gauge theory $\kappa$-symmetry conditions implies a specific map between the 
blackfold effective degrees of freedom (which control the form of the supergravity solution) and 
the abelian M5 brane theory degrees of freedom. This map is a central result of the paper. 

We conclude with a summary of the most pressing open issues and an outlook of the approach. 
Useful technical details are collected in two appendices.

\newsec{Abelian M5 brane theory}
\seclab\Mdbi

The theory of a single M5 brane is a six-dimensional abelian theory of the tensor multiplet with sixteen 
supersymmetries. The latter comprises of two complex Weyl spinors, a self-dual two-form potential $B_2$,
and five scalar fields. The dynamics of the long-wavelength fluctuations of the brane
is controlled by a non-linear effective action of these fields, which is the M5 brane analog of the 
Dirac-Born-Infeld action for D-branes. Following \PastiGX\ the bosonic part of this action (in short PST action) 
reads
\eqn\oneMaa{\eqalign{
S_{M5} = &- T_{M5} \int d^6\sigma \bigg( 
\sqrt{-\det (\gamma_{ab} +\tilde H_{ab})} + \frac{1}{4}\sqrt{-\det \gamma}
\, \tilde H^{ab}\HH_{abc} {\bf v}^c \bigg)
\cr
&+T_{M5} \int \bigg( C_6 +\frac{1}{2} \HH_3 \wedge C_3 \bigg)
~.}}
$T_{M5}=\frac{1}{(2\pi)^5\ell_P^6}$ is the M5 brane tension. The scalar fields $X^\mu$ (of which only 5 are
physical) define the induced worldvolume metric
\eqn\oneMab{
\gamma_{ab} = g_{\mu\nu} \d_a X^\mu \d_b X^\nu
}
where $a,b,\ldots$ are worldvolume indices and $g_{\mu\nu}$ the background metric. The $\HH_3$ field
strength is the gauge invariant 3-form
\eqn\oneMac{
\HH_3 = d B_2 - C_3
~.}
$C_3$ is the background supergravity 3-form potential and $C_6$ its Hodge dual. The self-duality of the
worldvolume 2-form potential is expressed nicely in the PST formulation with the use of an auxiliary 
scalar field $\varphi$ (usually referred to as $a$ in the literature). The derivatives of this field define the 
unit vector
\eqn\oneMad{
{\bf v}_a = \frac{\d_a \varphi}{\sqrt{-\d_b\varphi \, \d^b \varphi}}
}
which appears in \oneMaa. In \oneMaa\ we are also using the definitions
\eqn\oneMae{
\tilde H_{ab} = \HH^*_{abc} {\bf v}^c~, ~~
\HH^{*abc} = \frac{1}{3!} \frac{1}{\sqrt{-\det \gamma}} \varepsilon^{abcd_1d_2 d_3}\HH_{d_1 d_2 d_3}
~.}

Notice that the physical bosonic degrees of freedom of the tensor multiplet in six dimensions are 8:
5 from the scalars and 3 from the self-dual 2-form. The PST action is expressed in terms of $X^\mu$,
${\bf v}_a$, $\tilde H_{ab}$, which have in total 25 components.\foot{$X^\mu$ has 11 components, 
${\bf v}_a$ as a unit worldvolume vector has 5, and $\tilde H_{ab}$ has $15-6=9$ because of anti-symmetry 
and the defining relation ${\bf v}^a \tilde H_{ab}=0$. Several gauge invariances operate on this system.
For example, one can use a symmetry that shifts the auxiliary field $\varphi$
(and at the same time transforms the 2-form potential $B_2$) to fix the vector ${\bf v}_a$ and obtain a 
non-manifest Lorentz invariant formulation of the M5 brane effective theory \PerryMK.} 
In the supergravity analysis below we will also encounter a 25-component set of 
bosonic fields arising directly from the gravitational degrees of freedom.

The supersymmetric bosonic configurations of the M5 brane (which will be the main focus of this paper) 
obey the projection equation
\eqn\oneMaf{
\bG_\kappa\, \epsilon = \epsilon
}
where $\epsilon$ is a Majorana spinor in eleven-dimensions and $\bG_\kappa$
the M5 brane $\kappa$-symmetry matrix
\eqn\oneMag{
\bG_\kappa = 
\frac{{\bf v}_{a} t_{b}}{\sqrt{-\det(\gamma +\tilde H)}} \, \Gamma^{ab}
+\frac{\sqrt{-\det \gamma}  ~{\bf v}_{a}\tilde H_{bc}  }{2\sqrt{-\det(\gamma+\tilde H)}}\, \Gamma^{abc}
-\frac{\varepsilon_{a_1\ldots a_5b} {\bf v}^b {\bf v}_a}{5!\sqrt{-\det(\gamma+\tilde H)}} \, \Gamma^a 
\Gamma^{a_1\ldots a_5}
~.}
We defined 
\eqn\oneMai{
t^a = \frac{1}{8} \varepsilon^{ab_1b_2 c_1 c_2 d} \tilde H_{b_1 b_2} \tilde H_{c_1 c_2} {\bf v}_d
}
and $\Gamma^a$ are curved worldvolume $\Gamma$-matrices.

\subsec{1/2-BPS planar solution with uniform $\HH$-flux}
\subseclab\uniformflux

Before ending this section we summarize the properties of a 1/2-BPS solution that will soon play a 
protagonistic role in 
our discussion.

A simple solution of the PST equations of motion in flat space with trivial planar worldvolume geometry has 
(in the temporal gauge)
\eqn\oneMba{
{\bf v}^a= (1,0,0,0,0,0)
}
and constant $\HH$-flux with non-vanishing components\foot{For a general parametrization of constant
$\HH$-flux solutions see \BergshoeffJN. Such solutions give rise to a non-commutative M5 brane theory
and play a role in the general setup of OM theory \GopakumarEP.} 
\eqn\oneMbb{
\HH_{012} = \frac{H}{\sqrt{1+H^2}}~, ~~
\HH_{345} = 
H~, ~~ H={\rm constant}
~.}
For this profile $\tilde H_{12}=H$, the vector $t$ is vanishing and the $\kappa$-symmetry matrix is 
\eqn\oneMbc{
\bG_\kappa = - \frac{1}{\sqrt{1+H^2}} \left( H \Gamma_{||} + \Gamma_{||}\Gamma_\perp \right)
}
where $\Gamma_{||}=\Gamma^0 \Gamma^1 \Gamma^2$ and 
$\Gamma_\perp= \Gamma^3 \Gamma^4 \Gamma^5$.
Defining an angle $\theta$ such that\foot{In this definition $\theta\in \left [-\frac{\pi}{2},\frac{\pi}{2}\right]$.
With a parity transformation $\theta$ extends over the full range $[0,2\pi)$.} 
\eqn\oneMbd{
\cos\theta = \frac{1}{\sqrt{1+H^2}}~, ~~
\sin\theta = \frac{H}{\sqrt{1+H^2}}
}
the $\kappa$-symmetry equation \oneMaf\ becomes
\eqn\oneMbe{
\left(1+\sin\theta\, \Gamma_{||} +\cos\theta \, \Gamma_{||}\Gamma_\perp \right)\epsilon =0
~.}
This equation implies the reduction of the original thirty-two supersymmetries of the background by one half.

Since the $\HH$-flux induces M2 brane charge this solution is naturally interpreted as a 1/2-BPS state of the M5
with M2 brane charge along the directions (012) uniformly smeared along the transverse (345) plane
inside the M5 brane worldvolume.

\newsec{M5 branes in supergravity}
\seclab\Msugra

We now turn to the opposite regime where an infinite number of overlapping M5 branes is described by an 
extremal solution of the eleven-dimensional supergravity. In this section we fix our notation and present the 
symmetric solution whose long-wavelength deformations we will study later.

\subsec{Supergravity conventions, equations of motion, Killing spinor equations}

We shall use small Greek letters $\mu,\nu,\ldots$ to denote the curved spacetime indices and 
hatted small Greek letters $\hat \mu,\hat \nu,\ldots$ to denote tangent flat spacetime indices. 
Small latin letters $a,b,\ldots$ will be employed for spacetime directions parallel to the M5 brane 
worldvolume. The components of the metric, vielbein and spin connection are denoted respectively 
as $g_{\mu\nu}$, $e_\mu^{\hat \nu}$, $\omega_\mu^{~\hat \nu\hat \rho}$. Hodge duals in the 
eleven-dimensional spacetime will be written using a $\star$, and Hodge duals on effective worldvolumes
later using a $*$.

The bosonic part of the eleven-dimensional supergravity action is
\eqn\sugraaa{
I_{sugra}= \frac{1}{2\kappa_{11}^2}\int d^{11}x\, \sqrt{-g} \left( R-\frac{1}{2\cdot 4!}
F_{\mu_1\mu_2\mu_3\mu_4}F^{\mu_1\mu_2\mu_3\mu_4}\right)
-\frac{1}{12\kappa_{11}^2} \int C_3 \wedge F_4 \wedge F_4
}
where $\kappa_{11}$ is the eleven-dimensional Newton constant and 
$F_4=dC_3$ the four-form field strength. The equations of motion for the metric and gauge field are
\eqn\sugraab{
R_{\mu\nu}=\frac{1}{12}\left( (F_4^2)_{\mu\nu} -\frac{1}{12} g_{\mu\nu} F_4^2 \right)
~,}
\eqn\sugraac{
d \star F_4 +\frac{1}{2} F_4\wedge F_4=0
}
where the shorthand notation $F_4^2 = F_{\mu_1\mu_2\mu_3\mu_4}F^{\mu_1\mu_2\mu_3\mu_4}$
and $(F_4^2)_{\mu\nu} = F_{\mu\rho_1\rho_2\rho_3}F_\nu^{~ \rho_1\rho_2\rho_3}$ was employed.
In addition we have the Bianchi identity
\eqn\sugraad{
dF_4 =0
~.}

The supergravity multiplet also includes a single spin-$3/2$ field, the gravitino $\psi$.
Bosonic supersymmetric configurations with $\psi=0$ require the presence of a residual
supersymmetry expressed in terms of a Majorana spinor $\eta$ that obeys the Killing spinor equation
\eqn\sugraae{
\nabla_\mu \eta +\frac{1}{288} \left( \Gamma_\mu^{~\nu_1\nu_2\nu_3\nu_4} - 8 \delta_\mu^{~\nu_1}
\Gamma^{\nu_2\nu_3\nu_4} \right) F_{\nu_1\nu_2\nu_3\nu_4} \eta 
=0
~.}
We are using the standard notation where $\Gamma^{\mu\nu\ldots}$ denotes the antisymmetrized product 
of $\Gamma$-matrices
$\Gamma^\mu = e^\mu_{\hat \nu} \Gamma^{\hat \nu}$ obeying the Clifford algebra 
$\{ \Gamma^\mu ,\Gamma^\nu \}=2 g^{\mu\nu}$.

It is known \GauntlettFZ\ that by using the Killing spinor equation \sugraae\ on the identity 
\eqn\sugraaf{
\nabla_{[\rho}\nabla_{\mu]}\eta = \frac{1}{8} R_{\rho\mu\sigma_1\sigma_2} \Gamma^{\sigma_1\sigma_2}\eta
}
one can deduce the equation
\eqn\sugraag{\eqalign{
0 &= \left[ R_{\rho\mu} -\frac{1}{12} \left( (F_4^2)_{\rho\mu} -\frac{1}{12} g_{\rho\mu} F_4^2 \right)\right]
\Gamma^\mu \eta
\cr
&- \frac{1}{6 \cdot 3!} \star \left( d\star F_4 + \frac{1}{2} F_4 \wedge F_4 \right)_{\sigma_1\sigma_2\sigma_3}
\left( \Gamma_{\rho}^{~\sigma_1\sigma_2\sigma_3}
- 6\, \delta_\rho^{~\sigma_1}\Gamma^{\sigma_2\sigma_3}\right) \eta
\cr
& - \frac{1}{6!} (dF_4)_{\sigma_1\sigma_2\sigma_3\sigma_4\sigma_5}
\left( \Gamma_\rho^{~\sigma_1\sigma_2\sigma_3\sigma_4\sigma_5} - 10\, \delta_\rho^{~\sigma_1}
\Gamma^{\sigma_2\sigma_3\sigma_4\sigma_5} \right) \eta
~,}}
which implies, for example, that the Einstein equations follow automatically from the combination of the
Killing spinor equations, the Bianchi identity and the equations of motion of $F_4$.
This observation will be useful in the next section.

\subsec{Planar M2-M5 bound state solution}

The starting point of our discussion in the next section is an exact solution of the supergravity equations 
\sugraab-\sugraad\ that describes a planar 1/2-BPS M2-M5 bound state \Izquierdo\ 
(see also \refs{\HarmarkFF,\HarmarkRB})
\eqn\sugraba{
ds^2 = (HD)^{-\frac{1}{3}} \Big[ -(dx^0)^2 +(dx^1)^2 + (dx^2)^2 
+D \Big( (dx^3)^2 +(dx^4)^2 + (dx^5)^2 \Big)
+H\Big(dr^2 + r^2 d\Omega_4^2\Big)
\Big]
~,}
\eqn\sugrabb{
F_4 = d\CC_3 + D^{-1} \star d \CC_6
}
where
\eqn\sugrabc{
\CC_3 = -\sin\theta \, \left( H^{-1}-1 \right)\, dx^0 \wedge dx^1 \wedge dx^2 
+ \tan\theta \, DH^{-1}\, dx^3 \wedge dx^4 \wedge dx^5
~,}
\eqn\sugrabd{
\CC_6 = \cos\theta\, D \left( H^{-1} -1 \right) \, dx^0 \wedge dx^1 \wedge dx^2 \wedge dx^3 \wedge dx^4 
\wedge dx^5
~,}
\eqn\sugrabe{
H = 1+\frac{r_H^3}{r^3}~, ~~
D^{-1} = \cos^2\theta + \sin^2\theta\, H^{-1}
~.}
The solution is parametrized by two constants, $\theta$ and $r_H$, that control the energy density 
$\varepsilon$ and the M2 and M5 charge densities ($\QQ_2$, $\QQ_5$ respectively)
\eqn\sugrabf{
\varepsilon = \frac{\Omega_{(4)}}{16\pi G} r_H^3 ~, ~~
\QQ_2 = -\sin\theta \, Q~, ~~
\QQ_5 = \cos\theta \, Q~, ~~
Q= \frac{3 \Omega_{(4)}}{16\pi G} ~, ~~
8\pi G = \kappa_{11}^2
~.}
$\Omega_{(4)}= 8\pi^2/3$ is the volume of the unit round 4-sphere. 

The uniformly smeared M2-brane charge 
along the transverse (345) plane breaks the isotropy of the M5 brane worldvolume plane (012345) in direct
correspondence with the breaking that was noted in the non-gravitational constant $\HH$-flux solution of a 
single M5 brane \oneMba-\oneMbb\ in section \uniformflux.

\subsec{The Killing spinor equations of the planar M2-M5 bound state}
\subseclab\spinorzero

As warmup for the more complicated analysis that follows it is instructive to recall how the profile 
\sugraba-\sugrabe\ solves the Killing spinor equations \sugraae. 

The spacetime coordinates split naturally into the four groups: $(x^0,x^1,x^2)$, $(x^3,x^4,x^5)$, $r$ and
$(y^1, y^2, y^3, y^4)$ for the transverse $S^4$. Accordingly, we define the following antisymmetric 
combinations of the flat space $\Gamma$-matrices
\eqn\sugraca{
\Gamma_{||} = \Gamma^{\hat 0 \hat 1\hat 2}~, ~~
\Gamma_\perp = \Gamma^{\hat 3 \hat 4 \hat 5}~, ~~
\Gamma_\Omega = \Gamma^{\hat y^1 \hat y^2 \hat y^3 \hat y^4}
~.}
These combinations together with $\Gamma_{\hat r}$ obey the identity 
\eqn\sugracb{
\Gamma_{||}\Gamma_\perp \Gamma_\Omega \Gamma_{\hat r} = 1
~.}

For notational economy it is also convenient to define the functions
\eqn\sugracca{
e^S = HD~, ~~
e^Q = HD^{-2}~, ~~
e^R = r^{-6} H^{-2} D  
~,}
\eqn\sugraccb{
f_1 = -\sin\theta\, \left( H^{-1} - 1\right)~, ~~
f_2 = \tan\theta\, DH^{-1}~, ~~
f_3 = \cos\theta\, D\left( H^{-1}-1 \right)
~,}
which are all functions of the radial coordinate $r$.

With these specifications the covariant derivatives of the spinor $\eta$ are
\eqn\sugracda{
\nabla_{\mu} \eta = \d_{\mu}\eta +\frac{1}{12} \d^\nu S \, \Gamma_{\nu \mu}~, 
~~ \mu=0,1,2
~,}
\eqn\sugracdb{
\nabla_{\mu} \eta = \d_{\mu}\eta +\frac{1}{12} \d^\nu Q \, \Gamma_{\nu \mu}~, 
~~ \mu=3,4,5
~,}
\eqn\sugracdc{
\nabla_{y^i} \eta = \tilde \nabla_{y^i} \eta +\frac{1}{12} \d^\nu R \, \Gamma_{\nu y^i}~, 
~~ i=1,2,3,4
~,}
where $\tilde \nabla$ is the covariant derivative on the unit $S^4$.
In the background \sugraba-\sugrabe\ we take $\eta$ to be an $r$-dependent Killing spinor 
on $S^4$ \LuNU, namely we require
\eqn\sugrace{
\d_{\mu}\eta =0 ~~
(\mu=0,1,\ldots,5) ~, ~~
\tilde \nabla_{y^i} \eta = \frac{C}{2} e^{\frac{R}{6}} \Gamma_\Omega \Gamma_{y^i} \eta ~, ~~
C = \pm 1
~.}

The Killing spinor equation \sugraae\ can now be recast into the form
\eqn\sugracf{
\nabla_\mu \eta +\frac{1}{288} \left( -\frac{1}{2} \Gamma_\mu \Fslash 
+\frac{3}{2} \Fslash\, \Gamma_\mu \right)
\eta =0
}
where 
\eqn\sugracg{
\frac{1}{288} \Fslash = F_{||} + F_\perp + F_\Omega
~,} 
\eqn\sugraci{
F_{||} = \frac{1}{12} e^{\frac{S}{2}} \dslash f_1 \, \Gamma_{||}
~,~~
F_\perp = \frac{1}{12} e^{\frac{Q}{2}} \dslash f_2 \, \Gamma_\perp
~, ~~
F_\Omega = \frac{1}{12} r^{-2} e^{-\frac{R}{3}} D^{-1} \d_r f_3 \, \Gamma_\Omega
~.}
Writing out each of the components of \sugracf\ we obtain four equations
\eqn\sugracja{
\left( -\frac{1}{12} \dslash S + F_\perp +F_\Omega - 2 F_{||} \right) \eta =0
~,}
\eqn\sugracjb{
\left( -\frac{1}{12} \dslash Q + F_{||} +F_\Omega - 2 F_{\perp} \right) \eta =0
~,}
\eqn\sugracjc{
\d_r \eta - \Gamma_r \left( F_{||} +F_\perp -2 F_\Omega \right) \eta =0 
~,}
\eqn\sugracjd{
-\frac{C}{2} e^{\frac{R}{6}} \Gamma_\Omega \eta 
+ \left( -\frac{1}{12} \dslash R + F_{||} +F_\perp - 2 F_\Omega \right) \eta =0
~.}

By adding equations \sugracja, \sugracjb, \sugracjd, multiplying on the left by $\Gamma_r$, 
and using \sugracb\ we obtain the projection equation
\eqn\sugrack{
\Gamma_{||} \Gamma_\perp \eta = C\, \eta
~.}
Since both values of $C=\pm 1$ are allowed this equation is not restrictive for $\eta$.
With a similar manipulation of the linear combination \sugracja $+2 \times$\sugracjb\ 
(or equivalently from $2\times $ \sugracja\ $+$ \sugracjb)
we obtain a second projection equation
\eqn\sugracl{
\left( 1+ D^{\frac{1}{2}} H^{-\frac{1}{2}} \sin\theta \, \Gamma_{||} 
+D^{\frac{1}{2}} \cos\theta\, \Gamma_{||}\Gamma_\perp \right) \eta =0
~.}
Finally, \sugracjc\ combined with \sugracjd\ and \sugrack\ provides the differential equation
\eqn\sugracm{
\d_r \eta - \frac{1}{12} \d_r \log\left( H^{-2} D\right) \eta =0
~.}

The general solution of the system \sugracl-\sugracm\ is expressed in terms of a constant spinor
with sixteen independent components \Izquierdo, verifying the 1/2-BPS nature of the bound state.

\subsec{Comparison with the PST $\kappa$-symmetry equation}

Immediate and intuitive information about the number of preserved supersymmetries is obtained  
at the asymptotic infinity by analyzing 
the leading order form of the system \sugracl-\sugracm\ in a $1/r$-expansion around $r=\infty$. At leading order 
the spinor $\eta$ is constant, $\eta =\eta_0 + \OO(r^{-3})$, equation \sugracm\ is trivial and the projection 
equation \sugracl\ becomes
\eqn\sugradb{
\left ( 1+ \sin\theta\, \Gamma_{||} + \cos\theta\, \Gamma_{||}\Gamma_\perp \right) \eta_0 =0
~.}
The interesting, simple-minded observation is that this is the same as the $\kappa$-symmetry equation 
\oneMbe\ of a single M5 brane with the gauge-gravity identification of fields \oneMbd. 

The main purpose
of the ensuing sections is to exhibit how a similar analysis at infinity of more generic M5 brane configurations
produces a supergravity-induced $\kappa$-symmetry-type equation and how this compares with the 
original $\kappa$-symmetry equation of the abelian PST theory.

\newsec{Ansatz for extremal M2-M5 deformations}
\seclab\generalansatz

After a general $SO(1,5)$ rotation of the M5 brane worldvolume coordinates
\eqn\ansatzaa{
\sigma^a \to M^a_{~b}\, \sigma^b~, ~~ a=0,1,\ldots, 5~, ~~ M \in SO(1,5)
}
the solution \sugraba-\sugrabe\ takes the form
\eqn\ansatzaba{
ds^2 = \left(
e^{-\frac{S}{3}} \hat h_{ab} + e^{-\frac{Q}{3}} \hat \perp_{ab} \right) d\sigma^a d\sigma^b
+ e^{-\frac{R}{3}} (r^{-2} dr^2 + d\Omega_4^2)
~,}
\eqn\ansatzabb{
\CC_3 = - \sin\theta\, \left( H^{-1} -1 \right) \omega_3 - \tan\theta\, DH^{-1} *_6 \omega_3~, ~~
\CC_6 = \cos\theta\, D\left( H^{-1} - 1\right) \omega_6 
~.}
We used the $SO(1,5)$ matrix elements $M^a_{~b}$ to define three unit orthonormal vectors $u, v, w$
with components
\eqn\ansatzac{
u_a = M^0_{~a}~,  ~~ 
v_a = M^1_{~a}~, ~~
w_a = M^2_{~a}
~.}
This identification implies the orthonormality conditions 
(indices are lowered and raised with the six-dimensional Minkowski metric $\eta_{ab}$)
\eqn\ansatzad{
u_a u^a = -1~, ~~ v_a v^a = w_a w^a =1~, ~~ u_a v^a =0~, ~~ u_a w^a =0~, ~~ v_a w^a =0
~.}
The tensors appearing in \ansatzaba, \ansatzabb\ are expressed in terms of these vectors as follows
\eqn\ansatzaea{
\hat h_{ab} = - u_a u_b + v_a v_b + w_a w_b~, ~~ 
\hat \perp_{ab} = \eta_{ab} - \hat h_{ab}
~,}
\eqn\ansatzaeb{
\omega_3 = u\wedge v \wedge w~, ~~
\omega_6 = \sqrt{-\det\, \eta}~ d\sigma^0 \wedge d\sigma^1 \wedge d\sigma^2 \wedge d\sigma^3 
\wedge d\sigma^4 \wedge d\sigma^5 
~.}
The Hodge dual $*_6$ is taken with respect to the 6d metric $\eta_{ab}$.

The appearance of the vectors $u, v, w$ is a consequence of the breaking of the $SO(1,5)$ Lorentz symmetry 
induced by the presence of the smeared M2 brane charge. Accordingly, the tensors $\hat h_{ab}$ and 
$\omega_3$ are respectively a projector and a volume three-form along the M2 brane directions, and 
$\hat \perp_{ab}$ is the orthogonal projector.

Our goal is to determine the general supersymmetric deformations of the solution \ansatzaba, \ansatzabb\ 
in a long-wavelength expansion scheme. In the spirit of the fluid-gravity correspondence in 
AdS \BhattacharyyaJC, or the general blackfold approach \refs{\EmparanCS, \EmparanAT}, such a scheme 
arises essentially by promoting the parameters that control the zeroth order solution to slowly varying functions of 
the coordinates $\sigma^a$ and appropriately correcting the form of the solution order-by-order in the expansion 
to satisfy the supergravity equations of motion. The detailed construction of the perturbative solution requires 
the implementation of a technically complicated matched-asymptotic-expansion scheme where 
the supergravity equations are solved independently order-by-order in a near-zone and a far-zone region and 
then matched over a large intermediate region, called the overlap-zone (see \EmparanWM\ for a detailed
implementation of this scheme on black rings in neutral, pure Einstein gravity). In the case at hand, 
deformations with a characteristic long-wavelength scale $\RR \gg r_H$ define a near-zone that lies 
radially in the region $r \ll \RR$ and a far-zone that lies in the region $r \gg r_H$. 
$r_H$ is the scale appearing in \sugrabe. The large overlap region lies at distances $r$ such that 
$r_H \ll r \ll \RR$.

\subsec{Effective degrees of freedom}
\subseclab\dof

In this scheme the promoted parameters become naturally the degrees of freedom of an effective six-dimensional
worldvolume theory, which is thought to reside in the overlap zone. For deformations of the solution
\ansatzaba, \ansatzabb\ these parameters are
\eqn\ansatzba{
r_H~, ~~ \theta~, ~~ u~, ~~v~, ~~w~, ~~X^\mu
~.}
The scalars $X^\mu$ are degrees of freedom related to the breaking of the trasverse $SO(5)$ symmetry and
parametrize the promotion of the worldvolume metric $\eta_{ab}$ to the general induced metric
\eqn\ansatzbb{
\gamma_{ab} = g^{(0)}_{\mu\nu} \d_a X^\mu \d_b X^\nu
~.}
$g^{(0)}_{\mu\nu}$ is the asymptotic value of the bulk metric; here $g^{(0)}_{\mu\nu}=\eta_{\mu\nu}$.

In total, this effective worldvolume description of the gravitational dynamics gives rise to a formulation 
in terms of 25 degrees of freedom: 2 from $(r_H,\theta)$ that control the M2 and M5 charges, 
$3\times 6 - 3 \times 2 = 12$ from the unit orthonormal vectors $u,v,w$, and 11 from the scalars
$X^\mu$. As we noted above (see paragraph after eq.\ \oneMae), the same number of parameters
appears in the abelian PST effective action \oneMaa\ and $\kappa$-symmetry matrix \oneMag\ (including 
a vector built out of the auxiliary field $\varphi$ of PST that enforces the self-duality of the 2-form gauge field). 
Besides the transverse scalars that have an obviously common origin in both the gauge and gravity descriptions,
the rest of the degrees of freedom have a dramatically different looking form on each side.

\subsec{Near-zone supergravity deformation}
\subseclab\nearzone

We proceed to set up a specific ansatz for the first order deformation of the
supergravity fields. For the purposes of this paper, we will focus exclusively on the form of this ansatz 
in the near-zone region. Our proposal is motivated by analogous deformations of 
black brane solutions in pure Einstein gravity (see \refs{\EmparanWM,\CampsBR,\CampsHW}). 
It will be shown to be non-trivially consistent with known or expected properties.

Following \CampsHW\ (suitably extended to include all the fields of eleven-dimensional 
supergravity) we propose the following first order deformation of the bosonic fields \ansatzaba, \ansatzabb\
(\dd\ is a dummy variable that keeps track of the deformation order)
\eqn\ansatzcaa{
ds^2 = \left( e^{-\frac{S}{3}} \hat h_{ab} + e^{-\frac{Q}{3}} \hat \perp_{ab} \right) d\sigma^a d\sigma^b 
+ e^{-\frac{R}{3}} \left(r^{-2} dr^2 + d\Omega_4^2 \right)
+  \dde\, h_{\mu\nu}(x)\, dx^\mu dx^\nu
+ \OO(\dde^2)
~,}
\eqn\ansatzcab{
F_4 = d\CC_3 + D^{-1} \star d\CC_6 + \dde\, G_4 +\OO(\dde^2)
}
with
\eqn\ansatzcac{
\CC_3 = -\sin\theta\, \left( H^{-1} -1 \right) \omega_3 - \tan\theta \, DH^{-1} \star_{6} \omega_3~, ~~
\CC_6 = \cos\theta \, D\left( H^{-1}-1 \right) \omega_6
~.}

There are two new elements in these expressions, compared to the seed profile \ansatzaba, \ansatzabb. 
First, we have promoted all the previously constant parameters \ansatzba\
to $\sigma^a$-dependent fields. The functions $S, Q, R$ are still defined
in terms of $H, D$ as in \sugracca, \sugrabe, but they are now functions of both $r$ and $\sigma^a$ since 
$r_H = r_H(\sigma^a)$, $\theta = \theta(\sigma^a)$. The vectors $u, v, w$ are also functions of $\sigma^a$
on an effective curved worldvolume with induced metric $\gamma_{ab}$ \ansatzbb. 
The projectors $\hat h_{ab}$, $\hat \perp_{ab}$ and the forms $\omega_3$, $\omega_6$ are now
\eqn\ansatzcba{
\hat h_{ab} = - u_a u_b + v_a v_b + w_a w_b~, ~~ 
\hat \perp_{ab} = \gamma_{ab} - \hat h_{ab}
~,}
\eqn\ansatzcbb{
\omega_3 = u\wedge v \wedge w~, ~~
\omega_6 = \sqrt{-\det \gamma}~ d\sigma^0 \wedge d\sigma^1 \wedge d\sigma^2 \wedge 
d\sigma^3 \wedge d\sigma^4 \wedge  d\sigma^5 
}
and the Hodge dual $*_6$ is taken with respect to the 6d induced metric $\gamma_{ab}$.

The second modification includes the metric and 4-form corrections $h_{\mu\nu}$ and $G_4$. These are of the 
same order, $\OO(\dde)$, as the $\sigma$-derivatives of $\log r_H, \cos\theta$, $u,v,w$ and 
the derivatives of the velocities $\d_a X^\mu$. They need to be included in order to satisfy the supergravity 
equations at first order. 

As in Ref.\ \CampsHW\ the main strategy is to work locally around an arbitrary point on the effective worldvolume.
We assume that all the $\sigma^a$-dependent fields are slowly-varying fields of the worldvolume coordinates,
we expand them in a derivative expansion and work linearly in the perturbations. In this local linearization
of the perturbations, the fluctuations split naturally into two decoupled sets that can be analyzed independently.

The first set includes the {\it intrinsic} fluctuations, namely fluctuations that are neutral under the `R-symmetry'
generators of the $SO(5)$ that rotates the five-dimensional space transverse to the brane. These are fluctuations
of $r_H, \theta, u,v,w$ and the induced metric. The metric fluctuations are subleading to the first order that
we will be considering and can be neglected. 

The second set includes the ${\it extrinsic}$ fluctuations. These are fluctuations of the transverse scalars 
$X^\perp$, which, by definition, are charged under the transverse $SO(5)$.

The resulting perturbed bosonic fields are then inserted into the supergravity equations of motion 
\sugraab-\sugraad\ which are solved perturbatively to determine the deformed solution in the near-zone
region. 

At the level of supersymmetry the above deformations induce a corresponding perturbative expansion of the 
Killing spinor equations \sugraae. At first order the Majorana spinor $\eta$ is perturbed independently by the 
intrinsic and extrinsic fluctuations. One of the main goals of this paper is to exhibit the details of this perturbation.
The intrinsic perturbative Killing spinor equations will be discussed in section \intrinsic\ and the extrinsic ones in 
section \extrinsic.

\subsec{Bosonic blackfold equations}
\subseclab\bosequations

The implementation of the above scheme on the bosonic supergravity equations \sugraab-\sugraad\
produces a set of partial differential equations involving the effective degrees of freedom \ansatzba\ and
the field perturbations $h_{\mu\nu},G_4$ in \ansatzcaa-\ansatzcab. A subset of these equations
are {\it constraint} equations; they do not involve second derivatives of the radial coordinate, and do not 
involve the corrections $h_{\mu\nu},G_4$. These equations can be analyzed most easily in the 
asymptotic infinity of the overlap zone ($\RR \gg r\gg r_H$), where they yield a set of dynamical equations 
for the effective degrees of freedom \ansatzba\ only, the so-called {\it blackfold equations}. 

For the M2-M5 bound state in flat space the leading-order blackfold equations are\foot{This set of equations has 
already appeared in \refs{\NiarchosPN,\NiarchosIA}, where it was employed in the analysis of solutions 
that describe configurations of M2 branes ending orthogonally on M5 branes.
In \refs{\NiarchosPN,\NiarchosIA} the M2 brane current conservation equation $d(J_3+*_6 J_3)=0$
was mistakenly reported as $d*_6J_3=0$. However, for the solutions analyzed in \refs{\NiarchosPN,\NiarchosIA}
the missing term $dJ_3$ is automatically zero and does not affect the results.} (a supergravity derivation of these 
equations has been performed in \FluxBlackfolds)
\eqn\ansatzdaa{
D_a T^{ab} = 0~, ~~
d \left( J_3 + *_6 J_3\right) =0~, ~~
d J_6 =0
~,}
\eqn\ansatzdab{
K_{ab}^{~~\rho} T^{ab}=0
}
where $T^{ab}$, $J_3$ and $J_6$ are respectively the stress-energy tensor, M2 brane current, and M5 brane 
current of the effective worldvolume theory. $K_{ab}^{~~\rho}$ is the extrinsic curvature tensor for the induced
metric $\gamma_{ab}$ (see \EmparanAT\ for detailed expressions) and $D_a$ the worldvolume covariant 
derivative. All these quantities are functionals of the fields \ansatzba
\eqn\ansatzdba{
T^{ab} = \tilde Q\, r_H^3 \left( \sin^2\theta \, \hat h^{ab} + \cos^2\theta\, \gamma^{ab} \right)~, 
}
\eqn\ansatzdbb{
J_3 = -\tilde Q\, \sin\theta\, r_H^3 \, \omega_3~, ~~
J_6 = \tilde Q\, \cos\theta\, r_H^3 \, \omega_6~, ~~
\tilde Q =\frac{3\Omega_{(4)}}{4G}
~.}

The first set of equations \ansatzdaa\ comes from the analysis of the intrinsic fluctuations and the second 
\ansatzdab\ from the extrinsic. More specifically, the first equation in \ansatzdaa\ and equation 
\ansatzdab\ originate from the analysis of a particular combination of the metric equations \sugraab. 
Similarly, the conservation of the self-dual part of the current $J_3$ in \ansatzdaa\ comes from 
a component of the gauge-field equation \sugraac. The final equation in \ansatzdaa\ comes from the Bianchi 
identity \sugraad.

We notice that the six-current conservation, $dJ_6=0$, gives a simple constant of motion
\eqn\ansatzdc{
\cos\theta\, r_H^3 = {\rm constant}
~.}
Hence, when inserted into the remaining equations we obtain 
\eqn\ansatzdda{
D_a \left( \frac{\sin^2 \theta}{\cos\theta} \hat h^{ab} + \cos\theta \, \gamma^{ab} \right) =0
~,}
\eqn\ansatzddb{
\sin^2\theta\, \hat h^{ab} K_{ab}^{~~\rho} + \cos^2 \theta\, \gamma^{ab} K_{ab}^{~~\rho} =0
~,}
\eqn\ansatzddc{
d \left[ \tan\theta \left( \omega_3 + *_6\,  \omega_3 \right) \right] =0
~.}
This system of dynamical equations for the 25 effective worldvolume fields of the blackfold expansion
should be compared to the equations of motion of the PST action \oneMaa. Accumulating evidence
from explicit solutions of the DBI/PST theory and supergravity ($e.g.$ the BIon \refs{\GrignaniXM,\GrignaniXMM} 
and self-dual string soliton solutions \refs{\NiarchosPN}) 
suggests that the extremal equations of the supergravity/blackfold theory are equivalent 
to the equations of the abelian DBI/PST theory. The precise connection between the two, however, is not 
immediately obvious at the level of the bosonic equations.

The generic configuration obeying \ansatzdc-\ansatzddc\ is extremal but not supersymmetric. For example,
it can be an extremal, time-independent configuration (see \NiarchosIA\ for a stationary extremal configuration
of the M2-M5-KKW system). In order to determine the supersymmetric subset of 
solutions we need to analyze the supergravity Killing spinor equations \sugraae. Implementing 
the long-wavelength expansion of section \nearzone\ on the Killing spinor equations we will soon derive  
the supergravity analog of the $\kappa$-symmetry equation \oneMaf. Ultimately, this is expected to
allow us to determine which of the extremal blackfold configurations are supersymmetric and how many 
supersymmetries are preserved by the full supergravity solution. 

We note that according to the identity \sugraag\ the independent subset
of dynamical equations for supersymmetric configurations are, $e.g.$, the Killing spinor equations,
the equations of motion of the gauge field $C_3$ and its Bianchi identity. The statement that
appears to emerge out of this identity in the effective blackfold theory is that the set 
$\big \{$supergravity-induced $\kappa$-symmetry-like condition $\oplus$ charge current equations$\big \}$
implies the validity of the complete set of the bosonic blackfold equations 
\ansatzdaa, \ansatzdab. More specifically, since the equations in the above parenthesis are the
`constraint' part of the independent subset of supergravity equations, it is natural to anticipate that
by satisfying them we guarrantee also a solution of the full first order supergravity equations. 
Assuming this is correct, we deduce via \sugraag\ that the Einstein equations are also satisfied. 
That would imply that the corresponding extra constraint equations involving 
the effective stress-energy momentum conservation are also satisfied.

\newsec{Killing spinor equations: intrinsic deformations of M2-M5}
\seclab\intrinsic

The intrinsic perturbations are monopole deformations with respect to the transverse four-sphere.
We do not perturb the transverse scalars $X^\perp$, and working locally around a point (call it 
$\sigma=0$) in Riemann normal coordinates we can also take the induced metric to be flat
$\gamma_{ab}=\eta_{ab}$ \CampsHW. The expansion is organized by the number of worldvolume derivatives,
counted here by the power of the dummy variable $\dde$.
Working up to $\OO(\dde)$ we set
\eqn\intrinsicaa{
r_H(\sigma) = r_H(0) + \dde\, \sigma^a \d_a r_H +\OO(\dde^2)~,~~
\theta(\sigma) = \theta(0) +\dde\, \sigma^a \d_a \theta +\OO(\dde^2)
~,}
\eqn\intrinsicaba{
u^0(\sigma) = 1 + \OO(\dde^2)~, ~~
u^b(\sigma)  = \dde\, \sigma^a \d_a u^b ~, ~~ b =1,2,3,4,5
~,}
\eqn\intrinsicabb{
v^1(\sigma) = 1 + \OO(\dde^2)~, ~~
v^i(\sigma)  = \dde\, \sigma^a \d_a v^b ~, ~~ b =0,2,3,4,5
~,}
\eqn\intrinsicabc{
w^2(\sigma) = 1 + \OO(\dde^2)~, ~~
w^b(\sigma)  = \dde\, \sigma^a \d_a w^b ~, ~~ b =0,1,3,4,5
~.}
Three additional relations are satisfied by the first derivative corrections of the worldvolume vectors $u,v,w$
as a result of the orthonormality conditions
\eqn\intrinsicaca{
u^a v_a = 0 ~~\Rightarrow ~~ \d_a v_0 = \d_a u_1~, 
}
\eqn\intrinsicacb{
u^a w_a = 0 ~~\Rightarrow ~~ \d_a w_0 = \d_a u_2~, 
}
\eqn\intrinsicacc{
v^a w_a = 0 ~~\Rightarrow ~~ \d_a w_1 = - \d_a v_2
~.}
The corrections $h_{\mu\nu}$ and $G_4$ are $\sigma$-independent monopoles on the transverse sphere. 

These data fix the form of the first order perturbation of the bosonic fields \ansatzcaa, \ansatzcab, which we
repeat here for convenience
\eqn\intrinsicada{
ds^2 = \left( e^{-\frac{S}{3}} \hat h_{ab} + e^{-\frac{Q}{3}} \hat \perp_{ab} \right) d\sigma^a d\sigma^b 
+ e^{-\frac{R}{3}} \left(r^{-2} dr^2 + d\Omega_4^2\right)
+ \dde\, h_{\mu\nu} \, dx^\mu dx^\nu +\OO(\dde^2)
~,}
\eqn\intrinsicadb{
F_4 = d\CC_3 + D^{-1} \star d\CC_6 +\dde\, G_4 +\OO(\dde^2)
~,}
\eqn\intrinsicadc{
\CC_3 = -\sin\theta\, \left( H^{-1} -1 \right) \omega_3 - \tan\theta \, DH^{-1} \star_{6} \omega_3~, ~~
\CC_6 = \cos\theta \, D\left( H^{-1}-1 \right) \omega_6
~.}
Notice that these expressions contain implicit $\OO(\dde)$ contributions that arise from 
the $\sigma$-expansion of the functions $S,Q,R, \hat h_{ab}, \CC_3,\CC_6$.

The supersymmetric subset of these deformations obeys the Killing spinor equations \sugraae.
For such configurations the Killing spinor $\eta$, which at zeroth order is a function of the radial coordinate $r$ 
parametrized by the constants $r_H,\theta$, is perturbed accordingly to 
\eqn\intrinsicae{
\eta (r,\sigma,y^i) = \eta_0 +\dde\, \sigma^a \d_a \eta(r,0,y^i) +\dde\, \zeta(r,y^i) + \OO(\dde^2)
=  \chi (r,\sigma) \otimes \xi(y^i)
~.}
We used the notation $\eta_0 \equiv \eta(r,0,y^i)$ and $\zeta(r,y^i)$ is the fermionic analog of the corrections 
$h_{\mu\nu}$, $G_4$ that are needed to satisfy the full set of Killing spinor equations. $\xi$ is a Killing spinor
on the unit $S^4$ \sugrace\ with $C=\pm 1$.

A straightforward computation reveals that the $\OO(\dde)$ Killing spinor equations split into two groups. 
The first comes from the $\sigma$-independent part
\eqn\intrinsicafa{
\d_a \eta + \bar \Pi_1^{(a)} \, \eta_0 =0
~,}
\eqn\intrinsicafb{
\d_r \zeta + \bar \Pi_3\, \eta_0 + \Pi_4\, \zeta = 0~,
}
\eqn\intrinsicafc{
\left( C\,\bar \Pi_5^{(m)} + \bar \Pi_7^{(m)} \right) \eta_0 
 + \left( C\, \Pi_6^{(m)} + \Pi_8^{(m)}\right) \zeta = 0
~,}
where $m$ is an $S^4$ index.
The form of the operators $\bar \Pi_1^{(a)},\bar \Pi_3, \ldots$ is summarized in appendix \intKilling . 

The second group comes from the $\sigma$-linear piece of the Killing spinor equations
\eqn\intrinsicaia{
\Pi^{(b)}_{1,a} \, \eta_0 + \Pi_2^{(b)} \, \d_a \eta = 0
~, }
\eqn\intrinsicaib{
\d_r \d_a \eta + \Pi_{3,a} \, \eta_0 + \Pi_4 \, \d_a \eta =0
~,}
\eqn\intrinsicaic{
\left( C\, \Pi_{5,a}^{(m)} + \Pi_{7,a}^{(m)} \right) \eta_0 
 +\left(  C\, \Pi_6^{(m)} + \Pi_8^{(m)}\right) \d_a \eta = 0
~.}
The form of the operators $\Pi^{(b)}_{1,a}, \Pi_2^{(b)},\ldots$ is summarized in appendix \intKilling .
In contrast to the first group these equations do {\it not} involve the corrections $h_{\mu\nu}$, $G_4$, and 
therefore have the right features to play the Killing spinor counterpart of the bosonic constraint equations 
that give rise to the blackfold equations.

\subsec{$\kappa$-symmetry condition for intrinsic perturbations}
\subseclab\intrinsickappa

In section \spinorzero\ we described how the analysis of the $a=0,1,\ldots,5$ components of the supergravity
Killing spinor equation for the zeroth-order solution produces at the asymptotic infinity 
a $\kappa$-symmetry-like projection equation \sugradb\ that can be mapped to the abelian PST 
$\kappa$-symmetry equation \oneMbd, \oneMbe. For the perturbed system the analogous components
of the Killing spinor equation are \intrinsicafa, \intrinsicaia. We proceed to show how the $\sigma$-linear
subset \intrinsicaia\ produces a condition that will be mapped later to a perturbative version of 
the abelian PST $\kappa$-symmetry equation \oneMaf, \oneMag. We have checked that the remaining 
$\sigma$-linear equations \intrinsicaib, \intrinsicaic\ do not give rise to additional constraints on the Killing 
spinor at leading order in the $1/r$ expansion that we are considering.

In what follows it will be convenient to establish the notation $a_{||}$ for worldvolume indices in the range 
$(0,1,2)$ and $a_\perp$ for indices in the range $(3,4,5)$.

Expanding the operators $\Pi_2^{(b)}$, $\Pi_{1,a}^{(b)}$ around the asymptotic infinity in the overlap region
we find that the leading behavior is $\OO(r^{-4})$. Specifically, 
\eqn\intrinsicbaa{\eqalign{
\Pi_2^{(b_{||})} =& 
\frac{r_H^3(0)}{4r^4} \Gamma_{\hat b_{||}} \Gamma_{\hat r} 
\Big ( 1+ \sin^2 \theta(0)   
+ 2\sin\theta(0) \, \Gamma_{||}
\cr
&+\sin\theta(0) \cos\theta(0) \,  \Gamma_\perp
+\cos\theta(0)\, \Gamma_{\hat r} \Gamma_\Omega \Big) +\OO(r^{-7}) 
~,}}
\eqn\intrinsicbab{\eqalign{
\Pi_2^{(b_{\perp})} =& 
\frac{r_H^3(0)}{4r^4} \Gamma_{\hat b_{\perp}} \Gamma_{\hat r}
\Big ( 1-2 \sin^2 \theta(0)  
- \sin\theta(0) \, \Gamma_{||}
\cr
&-2\sin\theta(0) \cos\theta(0) \,  \Gamma_\perp
+\cos\theta(0)\, \Gamma_{\hat r} \Gamma_\Omega \Big) +\OO(r^{-7}) 
~,}}
and 
\eqn\intrinsicbd{\eqalign{
&\Pi_{1,a}^{(b_{||})}\eta_0 =
\frac{1}{4r^4} \delta_{b_{||}}^{\hat b_{||}} \Gamma_{\hat b_{||}} \Gamma_{\hat r} \times
\cr
&\Bigg\{
\d_a \left( (1+\sin^2\theta)r_H^3\right)  
+\d_a \left( \cos\theta \, r_H^3 \right) \Gamma_{||} \Gamma_\perp
+2\d_a \left( \sin\theta \, r_H^3 \right) \Gamma_{||}
+
\d_a \left( \sin\theta\cos\theta\, r_H^3 \right) \Gamma_\perp
\cr
&+\Big( 1+C \cos\theta(0)\Big) \Big( -2 + C \cos\theta(0)\Big) r_H^3(0) \times
\cr
&~~~~~~~~~~~~~~~~~~~~~~~~~~~~
\times \sum_{c_\perp} \delta^{c_\perp}_{\hat c_\perp} 
\left( -\d_a u_{c_\perp} \Gamma^{\hat 0 \hat c_\perp}
+\d_a v_{c_\perp} \Gamma^{\hat 1 \hat c_\perp}
+\d_a w_{c_\perp} \Gamma^{\hat 2 \hat c_\perp} \right) \Bigg\} \eta_0
\cr
&+\OO(r^{-7})
~,}}
\eqn\intrinsicbe{\eqalign{
&\Pi_{1,a}^{(b_\perp)} \eta_0 = 
\frac{1}{4r^4} \delta^{\hat b_\perp}_{b_\perp} \Gamma_{\hat b_\perp} \Gamma_{\hat r} \times
\cr
&\Bigg\{ 
\d_a \left( (1-2\sin^2\theta) r_H^3 \right)
+\d_a \left( \cos\theta \, r_H^3 \right) \Gamma_{||} \Gamma_\perp
-\d_a \left(\sin\theta \, r_H^3 \right) \Gamma_{||}
-2\d_a \left( \cos\theta \sin\theta \, r_H^3\right) \Gamma_\perp
\cr
&+\Big( 1+C \cos\theta(0)\Big) \Big( 1 -  2C \cos\theta(0)\Big) r_H^3(0)\times
\cr
&~~~~~~~~~~~~~~~~~~~~~~~~~~~~
\times \sum_{c_\perp} \delta^{c_\perp}_{\hat c_\perp} 
\left( -\d_a u_{c_\perp} \Gamma^{\hat 0 \hat c_\perp}
+\d_a v_{c_\perp} \Gamma^{\hat 1 \hat c_\perp}
+\d_a w_{c_\perp} \Gamma^{\hat 2 \hat c_\perp} \right) \Bigg\} \eta_0
\cr
&+\OO(r^{-7})
~.}}
Significant simplifications to the final expressions of $\Pi^{(b)}_{1,a}\, \eta_0$ were possible 
with the repeated use of the zeroth order equations \sugrack, \sugradb
\eqn\intrinsicbc{
\Gamma_{||}\Gamma_\perp\, \eta_0 = C\, \eta_0~, ~~
\left( 1+C  \cos\theta(0) +\sin\theta(0)\, \Gamma_{||} \right) \eta_0 = 0
~.}

We notice that the explicit dependence on the index $b_{||}$ and $b_\perp$ has disappeared in 
the quantities $\Gamma^{\hat b_{||}} \Pi_{1,a}^{(b_{||})}\, \eta_0$, 
$\Gamma^{\hat b_{\perp}} \Pi_{1,a}^{(b_{\perp})}\, \eta_0$.
Then, by taking, for example, the linear combination\foot{We remind that the same combinations were considered
at zeroth order.} 
$$
2 \Gamma_{\hat r} \Gamma^{\hat b_{||}} \times (\intrinsicaia ~with~b=b_{||}) ~+~
\Gamma_{\hat r} \Gamma^{\hat b_\perp} \times (\intrinsicaia ~with~ b=b_\perp)
$$
(and isolating the leading $\OO(r^{-4})$ terms) we discover the single independent condition from these 
equations
\eqn\intrinsicbf{
\Pi_{1,a} \, \eta_0 + \Pi_2 \, \d_a \eta =0
}
where 
\eqn\intrinsicbga{\eqalign{
&\Pi_{1,a} = 
\d_a r_H^3 
+\d_a \left( \cos\theta\, r_H^3 \right) \Gamma_{||} \Gamma_\perp
+ \d_a \left( \sin\theta\, r_H^3 \right) \Gamma_{||}
\cr
&- \Big( 1+C \cos\theta(0)\Big) r_H^3(0)
\sum_{c_\perp} \delta^{c_\perp}_{\hat c_\perp} 
\left( -\d_a u_{c_\perp} \Gamma^{\hat 0 \hat c_\perp}
+\d_a v_{c_\perp} \Gamma^{\hat 1 \hat c_\perp}
+\d_a w_{c_\perp} \Gamma^{\hat 2 \hat c_\perp} \right) 
~,}}
\eqn\intrinsicbgb{\eqalign{
\Pi_2 = r_H^3(0) \Big(
1+\sin\theta(0) \Gamma_{||} +\cos\theta(0) \Gamma_{||}\Gamma_\perp \Big)
~.}}
Recall that $\Pi_2\, \eta_0=0$ is the zeroth order Killing spinor equation \sugradb.
A simple consistency check of \intrinsicbf-\intrinsicbgb\ is performed in appendix \check.
We have also checked that these equations reproduce the expected 1/4-BPS supersymmetry of
the self-dual string soliton solution of Ref.\ \NiarchosPN.

The last central observation is that \intrinsicbf\ can be recast as a perturbative version of the abelian PST 
$\kappa$-symmetry equation \oneMaf, \oneMag\ with a specific mapping between gauge and gravity
variables. For starters, let us set 
\eqn\intrinsicbi{
\epsilon (\sigma) = f(\sigma) \, \eta(\sigma)
}
for the relation between the spinors appearing in equations \oneMaf\ and \intrinsicbf. The relative, 
generally $\sigma$-dependent, factor $f(\sigma)$ will be fixed in a moment. Expanding around $\sigma=0$ we 
then have locally
\eqn\intrinsicbj{
\epsilon(\sigma) = f(0)\, \eta_0 + \dde\, \sigma^a \left( f(0)\, \d_a \eta + \d_a f\, \eta_0 \right) +\OO(\dde^2)
~.}

Now consider an intrinsic fluctuation of the abelian PST $\kappa$-symmetry matrix around the zeroth order profile 
\oneMba, \oneMbb. In this deformation the originally constant fields ${\bf v}_a, \tilde H_{ab}$ become 
slowly-varying along the worldvolume, but the transverse scalars are kept constant and the worldvolume 
metric flat. Then, ${\bf \Gamma}_\kappa$ perturbs to 
\eqn\intrinsicbk{
{\bf \Gamma}_\kappa (\sigma) = {\bf \Gamma}_\kappa (0) +\dde\, \sigma^a \,{\bf \Gamma}_{\kappa,a} 
+ \OO(\dde^2)
}
where ${\bf \Gamma}_\kappa (0)$ is given by \oneMbc\ and 
\eqn\intrinsicbka{\eqalign{
{\bf \Gamma}_{\kappa,a} = \d_a {\bf \Gamma}_\kappa
=& \d_a \left( \frac{{\bf v}_{b} t_{c}}{\sqrt{-\det(\eta + \tilde H)}} \right) \Gamma^{bc}+
\cr
&+\d_a \left( \frac{{\bf v}_{b} \tilde H_{cd}}{2\sqrt{-\det (\eta +\tilde H)}} \right) \Gamma^{bcd}
-\d_a \left( \frac{1}{\sqrt{-\det (\eta +\tilde H)}} \right) \Gamma_{||}\Gamma_\perp
~.}}
Notice that there is no $\Gamma_\perp$ contribution in this expression. In addition, two possible
contributions to the last term from derivatives of the vector ${\bf v}$ cancel each other out.

Combining the expansions \intrinsicbj, \intrinsicbk, we find that the $\kappa$-symmetry equation \oneMaf\ 
is, by definition, satisfied at zeroth order. At first order we obtain 
\eqn\intrinsicbl{
\Big( f(0)\, \bG_{\kappa,a} + \left( \bG_\kappa(0) - 1 \right) \d_a f \Big) \eta_0 
+ f(0) \left( \bG_\kappa(0) -1 \right) \d_a \eta =0 
~.}

We conclude that a map between the perturbative field theory Killing spinor equation \intrinsicbl\ and 
the supergravity induced one \intrinsicbf\ is possible if we can set\foot{Since $\eta_0$ is a special spinor obeying
the zeroth order Killing spinor equation, it is important in the first equality to keep the action on it explicit.
Then, by using the zeroth order equations we can manipulate the form of the first order equations appropriately.}
\eqn\intrinsicbm{
\Pi_{1,a} \eta_0 =\Big( f(0)\, \bG_{\kappa,a} + \left( \bG_\kappa(0) - 1 \right) \d_a f \Big) \eta_0
~, ~~
\Pi_2 = f(0) \left( \bG_\kappa(0) -1 \right) 
~.}
One of the first checks is the absence of $\Gamma_\perp$ terms both in \intrinsicbf\ and \intrinsicbl.
Other terms, $e.g.$ terms proportional to the identity, also work properly. All potentially harmful terms that can 
spoil the match \intrinsicbm\ cancel out at the end of the computation.
Equations \intrinsicbm\ can be satisfied by requiring the gauge-gravity map
\eqn\intrinsicbna{
\d_a f = - \d_a \left( r_H^3 \right)
~,}
\eqn\intrinsicbnb{
\d_a \left( \frac{1}{\sqrt{-\det (\eta +\tilde H)}}  \right) = \d_a \left( \cos\theta \right)
~,}
\eqn\intrinsicbnc{
\d_a \left( -\frac{1}{3!} \varepsilon^{a_{||}b_{||}c_{||}} {\bf v}_{a_{||}} \tilde H_{b_{||}c_{||}} \right) 
= \d_a \left( \tan\theta(\sigma) \right)
~,}
\eqn\intrinsicbnd{
\sin\theta(0)\, \d_a {\bf v}_{c_\perp} 
+ \frac{1}{2} \varepsilon_{c_\perp}^{~~b_\perp d_\perp} \, \d_a \tilde H_{b_\perp d_\perp} = 
-\sin\theta(0)\, \d_a u_{c_\perp}
~,}
\eqn\intrinsicbne{
\d_a \tilde H_{2 c_\perp} = \tan\theta(0)\, \d_a v_{c_\perp}
~,}
\eqn\intrinsicbnf{
\d_a \tilde H_{1 c_\perp} = -\tan\theta(0)\, \d_a w_{c_\perp}
~.}

We observe that there are some field components on both sides that do not appear in this map, $i.e.$ are 
not needed in order to match the $\kappa$-symmetry conditions. On the abelian PST side these fields are
two of the $b_\perp$ components of $\d_a \tilde H_{0 b_\perp}$. On the supergravity/blackfold side 
the derivatives $\d_a u_{b_{||}}, \d_a v_{b_{||}}, \d_a w_{b_{||}}$ (of which only two are independent, see
\intrinsicaca-\intrinsicacc) do not appear. As we noted at the end of subsection \bosequations, however,
it is anticipated that the full set of bosonic blackfold equations follows from the 
combination of the Killing spinor equations {\it and} the charge current conservation equations. 
In general, the `missing' components will appear in these equations explicitly.
Analogous statements apply to the abelian PST side \Simon.

\newsec{Killing spinor equations: extrinsic deformations of M2-M5}
\seclab\extrinsic

A similar analysis can be performed for extrinsic deformations of the planar M2-M5 solution.
In this case the intrinsic variables ---$r_H, \theta$ and the vectors $u, v, w$ 
--- remain unperturbed. The deformation activates the 
transverse scalars $X^\perp$ and perturbs accordingly the induced worldvolume metric. 
As explained in detail in \CampsHW, it is convenient to work in a local adapted coordinate system employing
Fermi normal coordinates. This system assigns coordinates $(\sigma^a, z^i)$, $(i=6,\ldots,10)$, to the point
that lies a unit affine distance along the geodesic with tangent $\frac{\d}{\d z^i}$ orthogonally to the worldvolume
at $\sigma^a$. Then, perturbations around a locally flat worldvolume patch are induced by the extrinsic curvature
tensor $K_{ab}^{~~i}$ along each of the transverse directions $z^i$. The linear independence of these 
perturbations for each $i$ implies that we can set all but one to zero and study them independently. Accordingly,
we introduce a director cosine 
\eqn\extaa{
z^i = r \cos\phi
}
for a fixed $i$ and denote for convenience $K_{ab}^{~~i} \equiv K_{ab}$. Following \CampsHW\ we can then
bring the first order dipole deformation of the metric (in the near-zone region) into the form
\eqn\extaba{\eqalign{
ds^2 =& \left( e^{-\frac{S}{3}} \hat h_{ab} + e^{-\frac{Q}{3}} \hat \perp_{ab} - 2\,  \dde\, K_{ab}\, r\cos\phi \right)
d\sigma^a d\sigma^b 
\cr
&+e^{-\frac{R}{3}} \left( r^{-2} dr^2 + d\phi^2 +\sin^2\phi \, d\Omega_3^2 \right) 
+ \dde\, h_{\mu \nu}(r,\phi) \, dx^\mu dx^\nu +\OO(\dde^2)
~.}}
The projector $\hat h_{ab}$ is not perturbed. For convenience, in what follows we
set $\hat h = {\rm diag}(-1,1,1,0,0,0)$. The orthogonal projector is $\hat \perp_{ab} = \eta_{ab} - \hat h_{ab}$.
The functions $S$, $Q$, $R$ are not perturbed and are given by the zeroth order expressions \sugracca.

For the perturbed 4-form field strength we propose the ansatz
\eqn\extabb{
F_4 = d\CC_3 + D^{-1} \star d \CC_6 + \dde\, G_4(r,\phi) +\OO(\dde^2)
}
where again
\eqn\extabc{
\CC_3 = -\sin\theta \left( H^{-1} -1 \right) \omega_3 - \tan\theta \, D H^{-1} *_6 \omega_3 ~, ~~
\CC_6 = \cos\theta\, D \left( H^{-1}-1 \right) \omega_6
~.}
The form $\omega_3$ \ansatzcbb\ is not perturbed, but $\omega_6$ \ansatzcbb\ is perturbed in accordance 
with the worldvolume metric deformation
\eqn\extac{
\gamma_{ab} = \eta_{ab} - 2\, \dde\, K_{ab} \, r \, \cos\phi
~.}
Similarly, $*_6$ is perturbed according to \extac.

Compared to the previous section of intrinsic perturbations, now the bosonic corrections $h_{\mu\nu}$
and $G_4$ are dipole perturbations in the transverse sphere. 

For supersymmetric configurations the Killing spinor $\eta$ receives a corresponding dipole perturbation 
\eqn\extad{\eqalign{
\eta(r,\sigma,\phi, \vartheta^m) &= \eta_0 +\dde\, \eta_1 
\cr
&= \eta_0 +\dde\, \cos\phi \left( \lambda (r, \vartheta^m) +r\, K_{ab}(\sigma)\, \xi^{ab} (r,\vartheta^m) \right) 
+ \OO(\dde^2)
\cr
&=\eta_0 + \chi (r,\phi) \otimes \psi (\vartheta^m)
~.}}
$\vartheta^m$ ($m=1,2,3$) are coordinates on the unit $S^3$. $\eta_0$ is the zeroth order 
Killing spinor and $\eta_1$ its first order correction. In analogy to the previous discussion,
we have separated the contributions to $\eta_1$ into a piece 
$\xi^{ab}$ induced directly by the extrinsic curvature, and a second piece $\lambda$
needed to satisfy the full set of Killing spinor equations. $\lambda$ is also first order and 
proportional to $K_{ab}$ but does not vanish as $r \to 0$. Since we break the transverse $SO(5)$
symmetry, but retain an $SO(4)$ subset, we are now expressing the spinor correction $\eta_1$ 
as a tensor product with a unit $S^3$ Killing spinor $\psi$, whose covariant derivative on $S^3$ is by 
definition
\eqn\extae{
\nabla_m \psi = \frac{i\tilde C}{2} \Gamma_{m} \psi~, ~~ \tilde C =\pm 1
~.}
In this particular equation $\Gamma_m = e_m^{\hat m} \Gamma_{\hat m}$ where $e^{\hat m}_m$ is the
vielbein in the unit $S^3$. We note that $\eta_0$ is instead an $S^4$ Killing spinor with an 
a priori independent sign $C$ \sugrace.

Inserting this ansatz into the Killing spinor equations \sugraae\ we find the following set of perturbative
spinor equations
\eqn\extaea{
\Sigma_1^{(a)} \, \eta_0 + \Sigma_2^{(a)} \eta_1 =0
~,}
\eqn\extaeb{
\d_r \eta_1 + \Sigma_3 \, \eta_0 + \Sigma_4 \eta_1 =0
~,}
\eqn\extaec{
\d_\phi \eta_1 + \Sigma_5 \, \eta_0 + \Sigma_6\, \eta_1 =0
~,}
\eqn\extaed{
\Big( C\, \Sigma_7^{(m)}+\Sigma_9^{(m)} \Big) \eta_0 
+\Big( \tilde C\, \Sigma_8^{(m)} + \Sigma^{(m)}_{10}\Big) \eta_1 = 0
~.}
The more explicit form of the operators $\Sigma_i$ is summarized in appendix \extKilling.
When we implement on $\eta_1$ the ansatz of the second line in \extad\ we find 
as in section \intrinsic\ that the equations split into two groups, which are required to hold independently.
The first group does not receive any contributions from terms linear in $r\, K_{ab}$, but involves explicitly 
the corrections $h_{\mu\nu}$, $G_4$ and $\lambda$.
The second group depends linearly on $r\, K_{ab}$, but not on the corrections $h_{\mu\nu}$, $G_4$,
$\lambda$. We proceed to analyze some of the implications of the second group.

\subsec{$\kappa$-symmetry condition for extrinsic perturbations} 
\subseclab\kappaext

In direct analogy with the approach followed in section \intrinsic, we concentrate on the 
$(r\, K_{ab})$-dependent part of the first set of perturbative Killing spinor equations \extaea.
Expanding the operators $\Sigma_1^{(a)}$, $\Sigma_2^{(a)}$ around the asymptotic infinity in 
the overlap region we find the leading order equations 
\eqn\extba{
K_{ab} \left( \Sigma^{ab (c)}_1 \, \eta_0 + \Sigma_2^{(c)} \, \xi^{ab} \right) = 0
}
where 
\eqn\extbba{\eqalign{
&K_{bc} \, \Sigma^{bc(a_{||})}_1 \, \eta_0 =
\cos\phi\, \frac{r_H^3}{r^3}\,
\delta^{\hat a_{||}}_{a_{||}} \Gamma_{\hat a_{||}} \Gamma_{\hat r} 
\bigg[ -\frac{3}{2} \sin\theta\cos\theta\, \hat \Gamma
+\frac{1}{4} K \left( \sin\theta\cos\theta\, \Gamma_\perp - 
\cos\theta\, \Gamma_{||}\Gamma_\perp\right)
\cr
& -\frac{1}{4} \sin\theta\, \eta^{dd} K_{da} (\omega_3)_{dbc} \Gamma^{abc}
-\frac{1}{8} \sin\theta\cos\theta\, \eta^{dd} K_{da} (*_\eta \omega_3)_{dbc} \Gamma^{abc}
\bigg] \eta_0 + \OO(r^{-6})
~,}} 
\eqn\extbbb{\eqalign{
&K_{bc} \, \Sigma^{bc(a_{\perp})}_1 \, \eta_0 =
\cos\phi\, \frac{r_H^3}{r^3}\,
\delta^{\hat a_{\perp}}_{a_{\perp}} \Gamma_{\hat a_{\perp}} \Gamma_{\hat r} 
\bigg[ 3 \sin\theta\cos\theta\, \hat \Gamma
+\frac{1}{4} K \left(-2 \sin\theta\cos\theta\, \Gamma_\perp - 
\cos\theta\, \Gamma_{||}\Gamma_\perp\right)
\cr
& +\frac{1}{8} \sin\theta\, \eta^{dd} K_{da} (\omega_3)_{dbc} \Gamma^{abc}
+\frac{1}{4} \sin\theta\cos\theta\, \eta^{dd} K_{da} (*_\eta \omega_3)_{dbc} \Gamma^{abc}
\bigg] \eta_0 + \OO(r^{-6})
~,}} 
and
\eqn\extbbc{\eqalign{ 
\Sigma^{(c)}_2 = r\cos\phi\, \Pi_2^{(c)}
~.}}
$\Pi_2^{(c)}$ is the operator that appears already in equations \intrinsicbaa, \intrinsicbab. In \extbba, \extbbb\
we used the notation
\eqn\extbc{
K \equiv \eta^{ab} K_{ab}~, ~~ 
\hat \Gamma \equiv \frac{1}{3!} 
\varepsilon_{abcd_1 d_2 d_3}\, (\omega_3)_{e_1e_2 e_3}\, \eta^{d_1 e_1}\eta^{d_2 e_2} K^{d_3 e_3}\,
\Gamma^{abc}
~.}

Superficially, the leading order contributions to $K_{bc}\, \Sigma^{bc(a)}_1 \eta_0$ are order $\OO(r)$.
The total cancellation of these dangerous contributions is due to the identity (see appendix \Killidentity\ for an 
explicit derivation)
\eqn\extbd{
\left( \cos\phi\, \Gamma_{\hat r} - \sin\phi\, \Gamma_{\hat \phi} \right) \eta_0 =0
}
which follows essentially from the fact that $\eta_0$ is an $S^4$ Killing spinor. 
Further important cancellations
occur at the next order $\OO(r^{-3})$ because of \extbd\ and the zeroth order equation \sugradb.

As in section \intrinsickappa\ we consider the linear combination 
$$
2 \Gamma_{\hat r} \Gamma^{\hat a_{||}} \times (\extba ~with~a=a_{||}) ~+~
\Gamma_{\hat r} \Gamma^{\hat a_\perp} \times (\extba ~with~ a=a_\perp)
~.$$
Isolating the leading $\OO(r^{-3})$ terms we arrive at a single $(a_{||}, a_\perp)$-independent
equation of the form
\eqn\extbe{
K_{ab} \left( \Sigma_1^{ab} \, \eta_0 + \Pi_2 \, \xi^{ab} \right) =0
~,} 
with
\eqn\extbf{
K_{ab}\, \Sigma_1^{ab} =- r_H^3 \left(  K \cos\theta\, \Gamma_{||}\Gamma_\perp
+\frac{1}{2} \sin\theta\, \eta^{dd} K_{da} (\omega_3)_{dbc} \Gamma^{abc} \right) 
~,}
\eqn\extbg{
\Pi_2 = r_H^3 \left( 1+ \sin\theta\, \Gamma_{||} + \cos\theta\, \Gamma_{||} \Gamma_\perp \right) 
~.}

We can now show that this equation is the same as the $\kappa$-symmetry equation \oneMaf\
of the abelian PST theory perturbed around the constant-$\HH$ flux solution \oneMba, \oneMbb.
For extrinsic deformations (restricted to a specific transverse space direction $i$ for which $X_i=X$) 
the PST Killing spinor $\epsilon$ perturbs to 
\eqn\extbi{
\epsilon = \epsilon_0 + \dde\, X K_{ab} \, \epsilon^{ab} +\OO(\dde^2) 
= - r_H^3 \left( \eta_0 -\dde\, X K_{ab} \, \xi^{ab} +\OO(\dde^2) \right)
}
where in the second equality we used the map 
\eqn\extbia{
\epsilon_0 = - r_H^3\, \eta_0~, ~~
\epsilon^{ab} = r_H^3 \, \xi^{ab}
~.}
At the same time, the $\kappa$-symmetry matrix $\bG_\kappa$ \oneMag\ perturbs to
\eqn\extbj{
\bG_\kappa (X) = \bG_{\kappa}^{(0)} + \dde\, \delta \bG_\kappa + \ldots
~.}
The zeroth order term is 
$\bG_\kappa^{(0)}= - \left( \sin\theta\, \Gamma_{||} + \cos\theta\, \Gamma_{||} \Gamma_\perp \right)$
(see \oneMbc, \oneMbd). 
The first order term $\delta \bG_\kappa$ is induced by the metric perturbation 
\eqn\extbk{
\delta \gamma_{ab} = \gamma_{ab} -\eta_{ab} = -2 \, \dde\, X K_{ab} ~, 
}
at fixed `intrinsic' fields ${\bf v}_a$, $\tilde H_{ab}$. Using the profile of the zeroth order solution along with 
the definitions \oneMbd, and the variation identities
\eqn\extbm{\eqalign{
&\delta \left( \frac{1}{\sqrt{-\det (\gamma + \tilde H) }} \right) 
= -\frac{1}{2} \frac{1}{\sqrt{-\det (\gamma+ \tilde H)}} \gamma^{ab} \delta \gamma_{ab}
= \cos\theta \, X\, K
~,~~
\cr
&\delta \left( \frac{\sqrt{-\det \gamma}}{2\sqrt{-\det(\gamma+\tilde H)}} {\bf v}_a \tilde H_{bc} \,
e^a_{\hat a} e^b_{\hat b} e^c_{\hat c}\, \Gamma^{\hat a \hat b\hat c} \right)
=-\frac{1}{2} \sin\theta \, \eta^{dd} X K_{da} (\omega_3)_{dbc} \Gamma^{abc}
}}
we obtain 
\eqn\extbo{
\delta \bG_\kappa = 
-\frac{1}{2} \sin\theta \, \eta^{dd}  K_{da} (\omega_3)_{dbc} \Gamma^{abc}
- \cos\theta\, K\, \Gamma_{||}\Gamma_\perp
~.}
Assembling all the elements, the $\kappa$-symmetry equation \oneMaf\ becomes
\eqn\extbp{
r_H^3 \bigg[ -\left( K \cos\theta\, \Gamma_{||}\Gamma_\perp
+\frac{1}{2} \sin\theta \, \eta^{dd}  K_{da} (\omega_3)_{dbc} \Gamma^{abc} \right) \eta_0 
+ \left( 1+ \sin\theta\, \Gamma_{||} +\cos\theta\, \Gamma_{||}\Gamma_\perp \right)K_{ab} \xi^{ab} \bigg]
=0
}
reproducing the supergravity equations \extbe-\extbg.

\newsec{Open issues and outlook}

In this paper, following the lore of the blackfold approach \EmparanCS,
we have addressed the problem of long-wavelength supersymmetric deformations of M5 brane solutions
in eleven-dimensional supegravity. Initiating a study of the leading order perturbation
of the supergravity Killing spinor equations, we have shown that part of these equations
gives rise to a perturbative $\kappa$-symmetry-like condition for the blackfold effective worldvolume theory.
This equation exhibits the same structure as the $\kappa$-symmetry equation of the abelian
PST theory of a single M5 brane. Requiring a match between the two we have obtained a non-linear map
between the fields of the PST theory and the supergravity-derived fields of the blackfold effective 
theory. 

It would be very interesting to obtain a more covariant form of this map
extending the local analysis of this paper and including the charge current conservation equations.
This map would have several consequences. First, it would suggest an intriguing rewriting of the PST theory 
in a fluid-dynamical language. Could this lead to a fruitful reformulation of the theory on M5 branes?
Second, it would provide deeper insight into the blackfold equations. New supersymmetric 
solutions can be envisioned by converting the second order bosonic blackfold equations to first order
ones. Third, on a more conceptual level, this map would help elucidate a potential gauge-gravity 
equivalence for the full brane system in flat space.

The other important open problem is the full solution of the first order perturbed supergravity equations.
For that purpose one has to consider the complete set of Killing spinor equations in the near-zone, 
extend the analysis to the far-zone $(r \gg r_H)$ and finally perform the match in the overlap zone 
$(r_H \ll r \ll \RR)$.
This would establish a concrete relation between supersymmetric solutions of the leading order blackfold 
equations and full first-order corrected regular supergravity solutions. We anticipate this is a one-to-one
relation \EmparanCS. Progress in this problem may entail an educated use of the underlying $G$-structure
\GauntlettFZ\ of the seed solution and its deformation. This prospect is currently under investigation.
The higher orders of the expansion scheme are also of interest. The more constrained structures of 
supersymmetric solutions may lead in this context to a more tractable setup compared to the 
general non-supersymmetric, finite temperature situation.

Finally, although here we focused on M-theory and eleven-dimensional supergravity, it is natural to 
expect that analogous statements carry over to the brane solutions of other higher-dimensional supergravities. 
Branes in the ten-dimensional type IIA/B (and connections to the DBI theory) are an obvious context for future 
study.

\bigskip

\centerline{\bf Acknowledgements}

\bigskip
\noindent
I would like to thank C.\ Bachas, J.\ Camps, S.\ Katmadas, E.\ Kiritsis, N.\ Obers, G.\ Papadopoulos, 
A.\ Pedersen, K.\ Siampos, A.\ Tomasiello and A.\ Zaffaroni for enlightening discussions.
This work was supported in part by European Union's Seventh Framework Programme under grant 
agreements (FP7-REGPOT-2012-2013-1) no 316165, PIF-GA-2011-300984, the EU program ``Thales'' 
MIS 375734  and was also co-financed by the European Union (European Social Fund, ESF) and Greek national 
funds through the Operational Program ``Education and Lifelong Learning'' of the National Strategic Reference 
Framework (NSRF) under ``Funding of proposals that have received a positive evaluation in the 3rd and 4th Call 
of ERC Grant Schemes''.

\appendix{A}{Summary of perturbative Killing spinor equations}
\seclab\fullKilling

In this appendix we summarize the full set of intrinsic and extrinsic perturbations
of the Killing spinor equations \sugraae. We present the raw structure of the equations 
omitting details that are not used in the main text.

\subsec{Intrinsic perturbations}
\subseclab\intKilling

The intrinsic deformation of the metric $g_{\mu\nu}$ and four-form flux $F_4$ appears in 
eqs.\ \intrinsicada-\intrinsicadc. For the metric correction $h_{\mu\nu}$ it is convenient to choose
the gauge
\eqn\intKillaa{
h_{rr} =0~, ~~ h_{\mu y^i} = 0
}
where $y^i$ $(i=1,2,3,4)$ are coordinates of the transverse $S^4$. Then, \intrinsicada\ takes the
more specific form
\eqn\intKillab{\eqalign{
ds^2 &= \left( e^{-\frac{S}{3}} \hat h_{ab} + e^{-\frac{Q}{3}} \hat \perp_{ab} +\dde\, h_{ab} \right)
d \sigma^a d\sigma^b 
+2 \, \dde\, h_{ra} dr d\sigma^a 
\cr
&~~~~+ r^{-2} e^{-\frac{R}{3}} dr^2 
+e^{-\frac{R}{3}} \left( 1+ \dde\, h_\Omega \right) d\Omega_4^2 
+\OO(\dde^2)
\cr
& = ds_0^2 + \dde\, ds_1^2 +\OO(\dde^2)
~.}}
Do not confuse the projector $\hat h_{ab}$ with the metric perturbation $h_{ab}$.
The components $h_{ab}$, $h_{ra}$, $h_{\Omega}$ are functions of the radial coordinate $r$ only.
When the functions $S,Q,R,\hat h_{ab}$ are expanded in $\sigma$-derivatives the total $\OO(\dde)$ 
contribution is collected in the first order correction $ds_1^2$. Accordingly, the vielbein components
$e^{\hat \mu}_\nu$, and the components of the spin connection $\omega_\mu^{~\hat \nu\hat \rho}$ 
are shifted to
\eqn\intKillac{
e^{\hat \mu}_\nu = (e_0)^{\hat \mu}_\nu + \dde\, (e_1)^{\hat \mu}_\nu +\OO(\dde^2)
~,}
\eqn\intKillad{\eqalign{
\omega_\mu^{~\hat \nu\hat \rho} 
&= \eta^{\hat \rho \hat \mu} e_{\hat \mu}^\nu \d_{[\nu} e^{\hat \nu}_{\mu]}
- \eta^{\hat \nu \hat \mu} e^\nu_{\hat \mu} \d_{[\nu}e^{\hat \rho}_{\mu]}
+\eta^{\hat \rho \hat \mu} \eta^{\hat \nu \hat \sigma} \eta_{\hat \tau \hat \lambda} 
e^\nu_{\hat \mu} e^\sigma_{\hat\sigma} e^{\hat \tau}_\mu \d_{[\nu} e^{\hat \lambda}_{\sigma]}  
\cr
&= \left( \omega_\mu^{~\hat \nu\hat\rho} \right)_0 
+ \dde\, \left( \omega_\mu^{~\hat \nu\hat \rho} \right)_1 +\OO(\dde^2)
~.}}
We remind that the covariant derivatives of spinors are
\eqn\intKillada{
\nabla_\mu = \d_\mu +\frac{1}{4} \omega_\mu^{~\hat \nu \hat \rho}\, \Gamma_{\hat \nu\hat \rho}
=\d_\mu + \frac{1}{4} \left( \omega_\mu^{~\hat \nu\hat \rho} \right)_0\, \Gamma_{\hat \nu\hat \rho}
+ \frac{1}{4} \left( \omega_\mu^{~\hat \nu\hat \rho} \right)_1\, \Gamma_{\hat \nu\hat \rho} \, \dde +\OO(\dde^2)
~.}

After the implementation of the $\sigma$-expansion on the forms $\CC_3$ and $\CC_6$ in \intrinsicadb,
\intrinsicadc, the four-form flux $F_4$ expands similarly to 
\eqn\intKillae{
F_4 = (F_4)_0 + \dde\,  (F_4)_1 +\OO(\dde^2)
~.}
For the slash 
\eqn\intKillaf{
\Fslash = \Gamma^{\nu_1\nu_2\nu_3\nu_4} F_{\nu_1\nu_2\nu_3\nu_4}
}
we obtain
\eqn\intKillag{
\Fslash = \Fslash_0 +\dde\, \Fslash_1 +\OO(\dde^2)
}
where
\eqn\intKillai{
\Fslash_0 = \left( \PP_0 \right)^{\nu_1\nu_2\nu_3\nu_4}_{\hat\mu_1\hat\mu_2\hat\mu_3\hat\mu_4}\,
\Gamma^{\hat\mu_1\hat\mu_2\hat\mu_3\hat\mu_4} (F_{\nu_1\nu_2\nu_3\nu_4})_0
~,}
\eqn\intKillaj{
\Fslash_1 =  \left( \PP_0 \right)^{\nu_1\nu_2\nu_3\nu_4}_{\hat\mu_1\hat\mu_2\hat\mu_3\hat\mu_4}\,
\Gamma^{\hat\mu_1\hat\mu_2\hat\mu_3\hat\mu_4} (F_{\nu_1\nu_2\nu_3\nu_4})_1
+ 4  \left( \PP_1 \right)^{\nu_1\nu_2\nu_3\nu_4}_{\hat\mu_1\hat\mu_2\hat\mu_3\hat\mu_4}\,
\Gamma^{\hat\mu_1\hat\mu_2\hat\mu_3\hat\mu_4} (F_{\nu_1\nu_2\nu_3\nu_4})_0
~.}
We used the shorthand notation
\eqn\intKillak{
\left( \PP_0 \right)^{\nu_1\nu_2\nu_3\nu_4}_{\hat\mu_1\hat\mu_2\hat\mu_3\hat\mu_4}
\equiv \left( e_0 \right)^{\nu_1}_{\hat \mu_1} 
\left( e_0 \right)^{\nu_2}_{\hat \mu_2}
\left( e_0 \right)^{\nu_3}_{\hat \mu_3}
\left( e_0 \right)^{\nu_4}_{\hat \mu_4}   
~,}
\eqn\intKillal{
\left( \PP_1 \right)^{\nu_1\nu_2\nu_3\nu_4}_{\hat\mu_1\hat\mu_2\hat\mu_3\hat\mu_4}
\equiv \left( e_0 \right)^{\nu_1}_{\hat \mu_1} 
\left( e_0 \right)^{\nu_2}_{\hat \mu_2}
\left( e_0 \right)^{\nu_3}_{\hat \mu_3}
\left( e_1 \right)^{\nu_4}_{\hat \mu_4}   
~.}

These expressions, together with the expansion of the Killing spinor \intrinsicae
\eqn\intKillam{
\eta = \eta_0 +\dde\, \eta_1 + \OO(\dde^2)
~,}
are then inserted into the Killing spinor equations \sugraae
\eqn\intKillan{
\nabla_\mu \eta + \frac{1}{288} \left( -\frac{1}{2} \Gamma_\mu \Fslash +\frac{3}{2} \Fslash\, \Gamma_\mu\right) 
\eta =0
}
to obtain a set of $\OO(\dde)$ equations of the form
\eqn\intKillana{
\Pi_1^{(a)}\, \eta_0 + \Pi_2^{(a)} \, \eta_1 =0
~,}
\eqn\intKillanb{
\d_r \eta_1 + \Pi_3\, \eta_0 + \Pi_4 \, \eta_1 =0
~,}
\eqn\intKillanc{
\Big( C\, \Pi_{5}^{(m)} + \Pi_7^{(m)} \Big) \eta_0 
+\Big( C\, \Pi_6^{(m)}  + \Pi_8^{(m)} \Big) \eta_1 =0
~,}
where $m$ is an $S^4$ index and
\eqn\intKillaoa{\eqalign{
\Pi_1^{(a)}=&
\frac{1}{4} \left( \omega_a^{~~\hat \nu \hat \rho}\right)_1 \Gamma_{\hat \nu\hat \rho}
\cr
&+\frac{1}{288} \left(
-\frac{1}{2} \left( (e_0)^{\hat \mu}_a \Gamma_{\hat \mu} \Fslash_1 
+(e_1)^{\hat \mu}_a \Gamma_{\hat \mu} \Fslash_0 \right)
+\frac{3}{2} \left( \Fslash_0 (e_1)^{\hat \mu}_a \Gamma_{\hat \mu}
+\Fslash_1 (e_0)^{\hat \mu}_a \Gamma_{\hat \mu} \right) 
\right)
~,}}
\eqn\intKillaob{
\Pi_2^{(a)}=
\frac{1}{4} \left( \omega_a^{~~\hat \nu \hat \rho}\right)_0 \Gamma_{\hat \nu\hat \rho}
+\frac{1}{288} \left(
-\frac{1}{2} (e_0)^{\hat \mu}_a \Gamma_{\hat \mu} \Fslash_0
+\frac{3}{2} \Fslash_0 (e_0)^{\hat \mu}_a \Gamma_{\hat \mu}
\right)
~,}
\eqn\intKillaoc{\eqalign{
\Pi_3=&
\frac{1}{4}  \left( \omega_r^{~~\hat \nu \hat \rho}\right)_1 \Gamma_{\hat \nu\hat \rho}
\cr
&+\frac{1}{288}\left(
-\frac{1}{2} \left(  (e_0)^{\hat \mu}_r \Gamma_{\hat \mu} \Fslash_1 
+(e_1)^{\hat \mu}_r \Gamma_{\hat \mu} \Fslash_0 \right) 
+\frac{3}{2} \left( \Fslash_0 (e_1)^{\hat \mu}_r \Gamma_{\hat \mu} 
+\Fslash_1 (e_0)^{\hat \mu}_r \Gamma_{\hat \mu} \right)
\right)
~,}}
\eqn\intKillaod{
\Pi_4=
\frac{1}{4}  \left( \omega_r^{~~\hat \nu \hat \rho}\right)_0 \Gamma_{\hat \nu\hat \rho} 
+\frac{1}{288}\left(
-\frac{1}{2} (e_0)^{\hat \mu}_r \Gamma_{\hat \mu} \Fslash_0
+\frac{3}{2} \Fslash_0 (e_0)^{\hat \mu}_r \Gamma_{\hat \mu} 
\right)
~,}
\eqn\intKillaof{
\Pi_5^{(m)}= e^{\frac{R(0)}{6}}
\left( - h_{\Omega}\, (e_0)^{\hat m}_m
+ \frac{1}{6} \sigma^c \d_c R\, (e_0)^{\hat m}_m 
+ (e_1)^{\hat m}_m 
\right)
\Gamma_\Omega \Gamma_{\hat m} 
~,}
\eqn\intKillaoe{
\Pi_6^{(m)}=
e^{\frac{R(0)}{6}} (e_0)^{\hat m}_m \Gamma_\Omega \Gamma_{\hat m}
~,}
\eqn\intKillaog{\eqalign{
\Pi_7^{(m)}=&
\frac{1}{6} \d^r R(0)\, (e_0)^{\hat r}_r (e_1)^{\hat m}_m\, 
\Gamma_{\hat r} \Gamma_{\hat m}
\cr
&+\frac{1}{6} \Big(  \d^r R(0)\,  (e_1)^{\hat \mu}_r
+\sigma^c \d_c \d^r R\,  (e_0)^{\hat \mu}_r
+\d^a R \, (e_0)^{\hat \mu}_a 
\Big)
\Gamma_{\hat \mu} \Gamma_{\hat m} (e_0)^{\hat m}_m
\cr
&-\frac{1}{2} \d^r h_\Omega\, (e_0)_r^{\hat r} (e_0)^{\hat m}_m\,
\Gamma_{\hat r} \Gamma_{\hat m}
\cr
&+\frac{1}{288}\Big( -  (e_0)^{\hat m}_m \Gamma_{\hat m} \Fslash_1 
-(e_1)_m^{\hat m} \Gamma_{\hat m} \Fslash_0
+3 \left( \Fslash_0 (e_1)^{\hat m}_m \Gamma_{\hat m}  
+\Fslash_1 (e_0)_m^{\hat m} \Gamma_{\hat m} \right) 
\Big)
~,}}
\eqn\intKillaoh{
\Pi_8^{(m)}=
\frac{1}{6} \d^r R(0)\,  (e_0)^{\hat r}_r (e_0)^{\hat m}_m\, 
\Gamma_{\hat r} \Gamma_{\hat m}
+\frac{1}{288}\Big(- (e_0)^{\hat m}_m \Gamma_{\hat m} \Fslash_0
+3 \Fslash_0 (e_0)^{\hat m}_m \Gamma_{\hat m}
\Big)
~.}

By further implementing the ansatz \intrinsicae\ for $\eta_1$ and collecting the $\sigma$-linear part of
the operators $\Pi_{odd}$,
\eqn\intiKillap{
\Pi_{odd}(\sigma, r, y^i) = \bar \Pi_{odd}(r,y^i) +\ \sigma^a\, \Pi_{odd,a}(r,y^i)  
~,}
we find trivially that the equations \intKillana-\intKillanc\ split into the two independent groups
\intrinsicafa-\intrinsicafc\ and \intrinsicaia-\intrinsicaic.
In section \intrinsickappa\ we focused on the $\sigma$-linear part of the Killing spinor equations
\intKillana.

\subsec{Extrinsic perturbations}
\subseclab\extKilling

The analysis of the extrinsic perturbations proceeds in a similar fashion.
The metric deformation is now given by eq.\ \extaba. It is convenient to choose a gauge where
\eqn\extKillaa{
h_{\mu\nu}(r,\phi) dx^\mu dx^\nu = 
\cos\phi \left( \tilde h_{ab}(r) \, d\sigma^a  d\sigma^b 
+e^{-\frac{R}{3}} \left( \tilde h_r (r) \frac{dr^2}{r^2} +\tilde h_\Omega(r) \left( d\phi^2 +\sin^2\phi d\Omega_3^2\right)
\right) 
\right)
~.}
For further details about this choice we refer the reader to \CampsHW\ and references therein.
The four-form flux $F_4$ and the Killing spinor $\eta$ are perturbed as in \extabb, \extad.

A useful fact about covariant derivatives of spinors along the $S^3$ directions is that they can be written 
in terms of covariant derivatives on the unit $S^3$ as follows
\eqn\extKillaaa{\eqalign{
\nabla_m \eta = \nabla^{(S^3)}_m \eta - \frac{1}{4} \d^\nu
\left( \log \left( e^{-\frac{R}{3}} \sin^2\phi \left( 1+ \dde\, \tilde h_\Omega \, \cos\phi \right) \right) \right)
\Gamma_{\nu m} \eta
~.}}
Here $m$ is an $S^3$ index. Since $\eta_1$ is an $S^3$ Killing spinor \extae\ we find
\eqn\extKillaab{
\dde\, \nabla_m \eta_1 =
\frac{i \tilde C}{2\sin\phi} e^{\frac{R}{6}} \Gamma_m\, \dde\, \eta_1 
+ \frac{1}{2} e^{\frac{R}{3}} \left( \frac{1}{6} r^2  \d_r R\, \Gamma_{r m} 
- \Gamma_{\phi m} \right)\, \dde\, \eta_1
+\OO(\dde^2)
~.}
The covariant derivative of $\eta_0$, which is an $S^4$ Killing spinor, can be deduced using
the relation\foot{$\Gamma^{(S^4)}_\mu$ denotes a curved index $\Gamma$-matrix in $S^4$.}
\eqn\extKillaac{
\nabla_m^{(S^4)} \eta_0 = \nabla_m^{(S^3)} \eta_0 -\frac{1}{2} \cot\phi\, \Gamma^{(S^4)}_{\phi m} \eta_0
~.}
Together with the defining relation of $S^4$ Killing spinors we obtain
\eqn\extKillaad{
\nabla_m^{(S^3)} \eta_0 = \frac{1}{2} e^{\frac{R}{6}} \left( 1-\frac{1}{2}\, \dde\, \tilde h_\Omega \right)
\left( C\, \Gamma_\Omega \Gamma_m -\cot\phi\, \Gamma_{\hat \phi} \Gamma_m \right)\eta_0
~.}

Repeating the steps of the previous subsection \intKilling\ (appropriately adapted) we find that the 
Killing spinor equations \intKillan\ take the form
\eqn\extKillaba{
\Sigma_1^{(a)} \, \eta_0 + \Sigma_2^{(a)}\, \eta_1=0
~,}
\eqn\extKillabb{
\d_r \eta_1 + \Sigma_3 \, \eta_0 + \Sigma_4\, \eta_1 =0
~,}
\eqn\extKillabc{
\d_\phi \eta_1 + \Sigma_5\, \eta_0 + \Sigma_6 \, \eta_1 =0
~,}
\eqn\extKillabd{
\Big( C\, \Sigma_7^{(m)} + \Sigma_9^{(m)} \Big) \eta_0  
+\Big(\tilde C\, \Sigma_8^{(m)}  + \Sigma_{10}^{(m)} \Big) \eta_1 =0
~.}
$m$ is again an $S^3$ index. The general expression of the operators 
$\Sigma_1^{(a)}, \Sigma_2^{(a)}, \Sigma_3, \Sigma_4$ is the same as in \intKillaoa-\intKillaod.
The remaining operators are
\eqn\extKillaca{
\Sigma_5 = \frac{1}{4} \left( \omega_\phi^{~\hat \nu\hat \rho} \right)_1 \Gamma_{\hat \nu\hat\rho}
+\frac{1}{288}\left( -\frac{1}{2} \left(
(e_0)_\phi^{\hat \phi} \Gamma_{\hat \phi} \Fslash_1
+ (e_1)_\phi^{\hat \phi} \Gamma_{\hat \phi} \Fslash_0 
\right)
+\frac{3}{2} \left(
\Fslash_0 (e_1)_\phi^{\hat \phi} \Gamma_{\hat \phi} 
+ \Fslash_1 (e_0)_\phi^{\hat \phi} \Gamma_{\hat \phi} 
\right)
\right)
~,}
\eqn\extKillacb{
\Sigma_6 = \frac{1}{4} \left( \omega_\phi^{~\hat \nu\hat \rho} \right)_0 \Gamma_{\hat \nu\hat\rho}
+\frac{1}{288}\left( 
-\frac{1}{2} (e_0)_\phi^{\hat \phi}\, \Gamma_{\hat \phi} \Fslash_0
+\frac{3}{2} \Fslash_0 (e_0)_\phi^{\hat \phi}\, \Gamma_{\hat \phi} 
\right)
~,}
\eqn\extKillacc{
\Sigma_7^{(m)} = \frac{1}{2} e^{\frac{R}{6}} \Gamma_\Omega \Gamma_{\hat m}
\left( (e_1)^{\hat m}_m -\frac{1}{2} \tilde h_{\Omega} (e_0)^{\hat m}_m \right)
~,}
\eqn\extKillacd{
\Sigma_8^{(m)} = \frac{i}{2\sin\phi} e^{\frac{R}{6}} \Gamma_{\hat m} (e_0)^{\hat m}_m
~,}
\eqn\extKillace{\eqalign{
\Sigma_9^{(m)} =& 
-\frac{1}{2} e^{\frac{R}{6}} \cot\phi \, \Gamma_{\hat \phi} \Gamma_{\hat m}
\left( (e_1)^{\hat m}_m -\frac{1}{2} \tilde h_{\Omega} (e_0)^{\hat m}_m \right)
\cr
&-\frac{1}{4} \d_\nu \log\left( e^{-\frac{R}{3}}\sin^2\phi \right)
\left( (e_1)_{\hat \nu}^\nu (e_0)^{\hat m}_m +(e_0)_{\hat \nu}^\nu (e_1)^{\hat m}_m \right)
\Gamma^{\hat \nu} \Gamma_{\hat m}
\cr
&-\frac{1}{4} \d_\nu \left( \tilde h_\Omega \cos\phi \right) (e_0)_{\hat \nu}^\nu (e_0)^{\hat m}_m\,
\Gamma^{\hat \nu} \Gamma_{\hat \mu}
\cr
&+\frac{1}{288}\left( -\frac{1}{2} \left( (e_0)_m^{\hat m} \Gamma_{\hat m} \Fslash_1
+ (e_1)_m^{\hat m} \Gamma_{\hat m} \Fslash_0 \right)
+\frac{3}{2} \left( \Fslash_0 (e_1)_m^{\hat m} \Gamma_{\hat m} 
+ \Fslash_1 (e_0)_m^{\hat m} \Gamma_{\hat m} \right)
\right)
~,}}
\eqn\extKillaca{
\Sigma_{10}^{(m)} = 
\frac{1}{2} e^{\frac{R}{6}} \left( \frac{1}{6} r\d_r R \, \Gamma_{\hat r} - \Gamma_{\hat \phi} \right) 
\Gamma_{\hat m} (e_0)^{\hat m}_m
+\frac{1}{288}\left( 
-\frac{1}{2} (e_0)_m^{\hat m} \Gamma_{\hat m} \Fslash_0
+\frac{3}{2} \Fslash_0 (e_0)_m^{\hat m} \Gamma_{\hat m} 
\right)
~.}

Implementing the ansatz \extad\ and collecting separately the pieces that depend linearly on 
$r\, K_{ab}$ we find that eqs.\ \extKillaba-\extKillabd\ split into two independent
groups. In section \kappaext\ we focused on a specific part, \extba, of the $(r\, K_{ab})$-linear group.

\appendix{B}{Useful identities}

\subsec{Consistency check of \intrinsicbf}
\subseclab\check

The pertubative Killing spinor equations \intrinsicbf\ imply
\eqn\checkaa{
\Pi_2\, \d_a \eta = -\Pi_{1,a}\, \eta_0 ~~\Rightarrow ~~
\left( \frac{1}{2r_H^3} \Pi_2 - 1\right)  \Pi_{1,a} \, \eta_0=0
}
where we multiplied simultaneously both sides by $\Pi_2$, used the identity
\eqn\checkab{
\Pi_2 ^2 = 2r_H^3 \Pi_2
~,}
and re-applied the equation \intrinsicbf. In this appendix
we examine the validity of the consistency equation \checkaa.

For convenience let us write
\eqn\checkac{
\Pi_{1,a} = Q_a +R_a
}
with 
\eqn\checkad{
Q_a = \d_a r_H^3 + \d_a (\sin\theta r_H^3) \Gamma_{||} + \d_a (\cos\theta r_H^3) \Gamma_{||}\Gamma_\perp
~,}
\eqn\checkae{
R_a = -(1+C\cos\theta)\, r_H^3(0)\,
\sum_{c_\perp} \left( -\d_a u_{c_\perp} \Gamma^{\hat 0 \hat c_{\perp}}
+\d_a v_{c_\perp} \Gamma^{\hat 1 \hat c_{\perp}}
+\d_a w_{c_\perp} \Gamma^{\hat 2 \hat c_{\perp}}
\right)
~.}
With a bit of algebra one can show that 
\eqn\checkaf{
\left( \frac{1}{2r_H^3} \Pi_2 - 1\right) Q_a\, \eta_0
= \frac{C \sin\theta\, \d_a \theta}{2(1+C\cos\theta)} \Pi_2\, \eta_0 =0
} 
and
\eqn\checkag{\eqalign{
\left( \frac{1}{2r_H^3} \Pi_2 - 1\right) R_a\, \eta_0
= \frac{1}{2} \sin\theta   
\sum_{c_\perp} \left( -\d_b u_{c_\perp} \Gamma^{\hat 0 \hat c_{\perp}}
+\d_b v_{c_\perp} \Gamma^{\hat 1 \hat c_{\perp}}
+\d_b w_{c_\perp} \Gamma^{\hat 2 \hat c_{\perp}} \right)
\Gamma_{||} \Pi_2\, \eta_0 =0
~.}}
We conclude that the consistency equation \checkaa\ is satisfied automatically as it should.

\subsec{An identity for Killing spinors on $S^4$}
\subseclab\Killidentity

In this appendix we prove the identity \extbd\ that was employed in section \kappaext.
Following section \spinorzero\ we consider a spinor $\eta$ in eleven dimensions 
whose four-sphere part is a Killing spinor on the unit $S^4$. By definition, the covariant derivatives
of $\eta$ on $S^4$ obeys the identity
\eqn\idaa{
\nabla_j \eta = \frac{C}{2} \Gamma_\Omega \Gamma_{j} \, \eta
~, ~~ C=\pm 1
~.}
$\Gamma_\Omega$ is the chirality operator on $S^4$ and $\Gamma_j = e_j^{\hat j} \Gamma_{\hat j}$
with $e_j^{\hat j}$ the vielbein on the unit $S^4$. In addition, we require the identity \sugrack, 
equivalently
\eqn\idaaa{
\Gamma_{\hat r} \Gamma_\Omega \, \eta = C\, \eta
~.}

In hyperspherical coordinates, where the metric of the unit four-sphere is
\eqn\idab{
d\Omega_4^2 = d\phi^2 +\sin^2\phi \, d\Omega_3^2
~,}
$\eta$ takes the form \LuNU
\eqn\idac{
\eta = e^{\frac{C}{2} \phi \Gamma_\Omega \Gamma_{\hat \phi}} \tilde \eta~, ~~
{\rm such~that}~~\d_\phi \tilde \eta =0
~.}
A convenient standard identity of $\Gamma$-matrix exponentials reads
\eqn\idad{
e^{\frac{C}{2} \phi \Gamma_\Omega \Gamma_{\hat \phi}} 
= \cos\left( \frac{\phi}{2}\right) \cdot {\bf 1} 
+ C \sin\left( \frac{\phi}{2} \right) \Gamma_\Omega \Gamma_{\hat \phi}
~.}

Combining \idaaa\ with the more explicit form \idac\ it is straightforward to show that
\eqn\idae{
\left[ \cos\left( \frac{\phi}{2} \right) \Gamma_{\hat r} - \sin\left( \frac{\phi}{2} \right) \Gamma_{\hat \phi}\right] 
\eta =0
~.}
Applying this equation at angle $2\phi$ and using 
$\eta(2\phi) = e^{\frac{C}{2}\phi \Gamma_\Omega\Gamma_{\hat \phi}} \eta(\phi)$
we arrive easily at the required identity \extbd
\eqn\idaf{
\left( \cos\phi \, \Gamma_{\hat r} - \sin\phi\, \Gamma_{\hat \phi} \right) \eta =0
~.}

\listrefs
\end